\documentclass[reprint,amsmath,amssymb,aps]{revtex4-2}
\usepackage{graphicx}
\usepackage{dcolumn}
\usepackage{bm}
\usepackage{wrapfig}
\usepackage{soul,xcolor}
\usepackage{mathtools}
\usepackage{array}
\usepackage{comment}
\usepackage{makecell}
\usepackage{float}
\begin{document}

\title{Effect of (001) and (111) epitaxial strain on \emph{Pnma} Perovskite oxides}

\author{Amartyajyoti Saha,$^{1,2}$ Turan Birol$^1$}
\affiliation{$^1$Department of Chemical Engineering and Materials Science, University of Minnesota}
\affiliation{$^2$School of Physics and Astronomy, University of Minnesota}

\begin{abstract}

With recent advances in strain engineering and its widespread applications, it is becoming increasingly important to understand the effect of biaxial strain on the most common structural phase of perovskites -- the orthorhombic $Pnma$ structure. 
In this work, by using a combination of group theory and first principles density functional theory, we study the effect of biaxial strain on (001) and (111)-oriented CaTiO$_3$, SrSnO$_3$ and SrZrO$_3$ films.
We observe manifestly different behaviors depending on the strain orientation, with common trends emerging between different materials. In addition to many different structural phases observed in individual compounds before, we identify a transition between two different phases with the same space group name ($P2_1/c$) but different symmetries in (111)-strained materials. 
We also find that allowing the relaxation of the out-of-plane monoclinic angles, often ignored in first principles studies, lead to significant stabilization of certain phases and is essential to correctly determine the structural ground state.
\end{abstract}

\maketitle

\section{Introduction}

Biaxial strain, applied on thin films by growing them on lattice mismatched substrates, is a commonly used strategy to modify properties of perovskites and related materials \cite{Moloney2020,Schlom2014}. 
For example, strain engineering has significant potential to improve the performance and stability of halide perovskite solar cells \cite{Moloney2020,Zhu2019,Jiao2021}. It has  also been used to tune the emission wavelength of perovskite-based light emitting diodes \cite{Liu2022, Yang2020}. 
A large body of strain engineering work is also focused on perovskite oxide thin films, and their  dielectric and ferroelectric properties \cite{Dimos1998, Schlom2007, Schlom2007, Martin2017, Fernandez2022}.

Most perovskite oxides have one more more structural distortions that make them deviate from the cubic high-symmetry reference structure. By far the most common distortion is the oxygen octahedral rotations that emerge due to the small size and hence under-bonding of the $A$-site cations in $AB$O$_3$ \cite{Lufaso2001, Woodward1997b}. The octahedral rotations around different cubic axes can have different magnitudes and patterns, which leads to many different space group symmetries \cite{Woodward1997a, Glazer1972, Howard1998}. The octahedral rotations can also couple with and give rise to other interesting phenomena, such as metal insulator transitions or improper ferroelectricity, which can often be controlled by strain since octahedral rotations often couple with strain strongly \cite{Wang2022RTO, Najev2022, Choquette2016}. 

More than half of all oxide perovskites have the same oxygen octahedral rotation pattern, shown as $a^-a^-c^+$ in Glazer notation \cite{Glazer1972}, which leads to the orthorhombic space group $Pnma$ also observed in the prototypical CaTiO$_3$ \cite{Kay1957}. When under strain, different $Pnma$ perovskites stabilize different phases. For example CaTiO$_3$ attains a polar space group \cite{Biegalski2015}, whereas SrSnO$_3$ is shown to stabilize the tetragonal $I4/mcm$ phase under compressive strain \cite{Wang2018}. While there are a large number of first-principles  studies on individual compounds, an overarching study of biaxially strained $Pnma$ perovskites that compares the trends in different compounds and different strain directions is missing to date to the best of our knowledge. In particular, most of the studies on biaxially strained perovskites focus on (001) films, and first principles work on (111) $Pnma$ perovskites is not common. This is despite the fact that there are many exciting theoretical predictions in these systems \cite{Okamoto2013, Ruegg2011} and a growing number of (111)-oriented thin films grown by MBE or pulsed laser deposition; for example, SrFeO$_{2.5}$ \cite{Chakraverty2010}, PbTiO$_3$ \cite{Tang2019PTO}, SrTiO$_3$ \cite{Liang2015}, SrVO$_3$ \cite{Roth2021SVO111}, various nickelates \cite{Middey2014, Middey2016, Gibert2016} and most recently KTaO$_3$ \cite{Kim2023KTO} have been grown with (111) orientation. 

In this work, we use first principles density functional theory to study the structural evolution of three different $Pnma$ perovskites under biaxial strain. 
As three distinct examples,  we consider CaTiO$_3$ \cite{Kay1957}, SrZrO$_3$ \cite{Roosmalen1992} and SrSnO$_3$ \cite{Glerup2005} with Goldschmidt tolerance factors 0.846, 0.861, and 0.873 respectively. Despite having very different $B$-site cations, all three of these perovskites have the $Pnma$ symmetry at room temperature, and undergo a series of structural phase transitions to become cubic ($Pm\bar{3}m$) at high temperature \cite{Yashima2009, Glerup2005, Kennedy1999}. 
In order to capture different symmetry states induced by different strain directions, we go beyond the commonly studied (001) films by considering (111) films as well. By considering many possible rotation patterns and structural phases, and projecting the structures obtained from DFT on irreducible representations (irreps) of the space group, we obtain some general trends regarding the effect of strain on the crystal structure in $Pnma$ perovskites. Some of our results include that \textit{(i)} compressive (001) strain increases the octahedral rotation tendencies around the out-of-plane strongly, whereas tensile strain has a relatively mild effect on rotations around in-plane axes, \textit{(ii)} as a result of this and the competition between the ferroelectric polarization and octahedral rotations, tensile (001) strain is more conducive to ferroelectricity than compressive strain, and \textit{(iii)} (111) thin films don't develop ferroelectric instabilities and phases as easily as the (001) films, since the shear-like strain due to the expansion or compression of the (111) planes is more favorably accommodated by changing the out-of-plane spacing of atomic layers, rather than the emergence of a polarization. We also show \textit{(iv)} the presence of a transition between two nonpolar phases with different symmetry but the same space group name under (111) strained films, and that \textit{(v)} relaxing the monoclinic angle between the out-of-plane axis and the in-plane ones in biaxial relaxations can be important in finding the correct structural ground state. 

This paper is organized as follows: In Sec.~\ref{sec:methods}, we introduce the details of our computational methods. In Sec.~\ref{sub:instability}, we discuss the common structural instabilities in the $Pnma$ structure. In Sec.~\ref{sub:001} and Sec.~\ref{sub:111} , we discuss the relative stability of different structures and the effect of biaxial strain on the rotation irreps under (001) and (111) strain, respectively. We compare the similarities and differences between the (001) and (111) strain in  Sec.~\ref{sub:discussion}. We conclude with a summary in Sec.~\ref{sec:conclusion}.

\section{\label{sec:methods}Methods}
\begin{figure}[t]
    \centering
    \includegraphics[width=\linewidth]{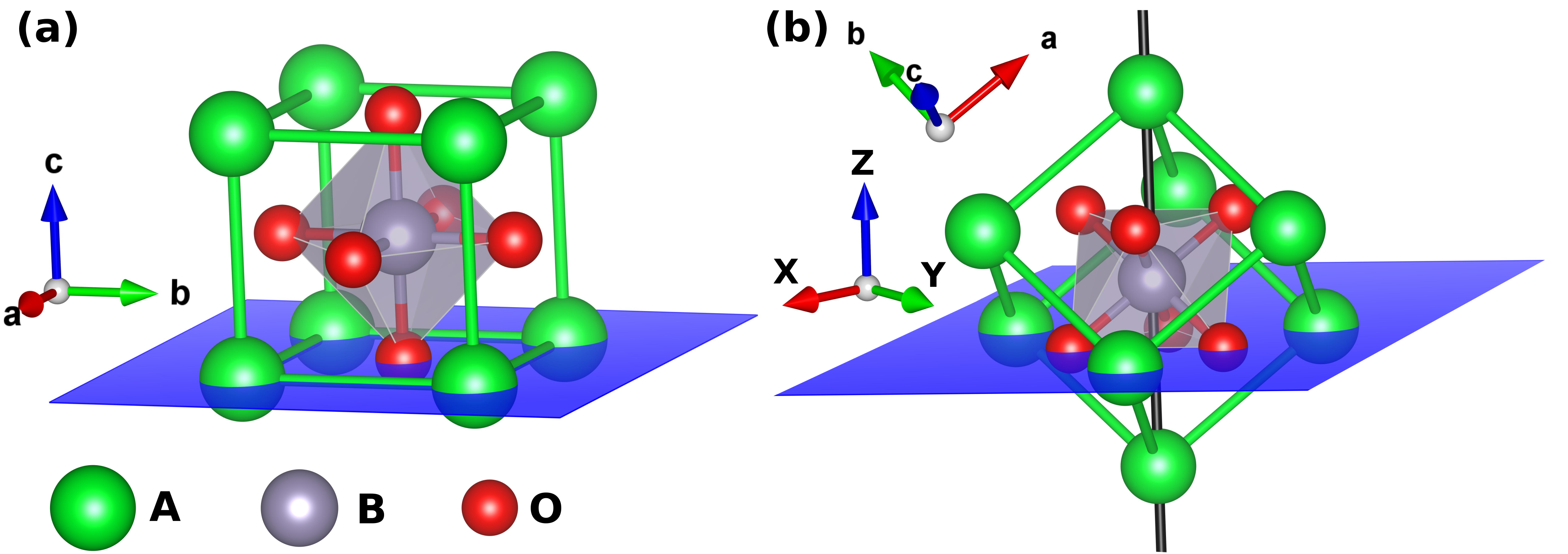}
    \caption{(a) Cubic perovskite structure under (001) epitaxial strain. We relax the magnitude and direction of $c$-axis while keeping the $ab$ plane on a rigid square grid. (b) Cubic perovskite structure under (111) epitaxial strain. The $Z$-axis is parallel to the black line through the B site. We fix the structure on a rigid square grid along $XY$ plane and relax the magnitude and direction of $Z$-axis.}
    \label{fig:Strain}
\end{figure}

First-principles density functional theory calculations were performed with the revised Perdew-Burke-Ernzerhof generalized gradient approximation for solids (PBEsol) using the Vienna \textit{ab initio} simulation package (\textsc{vasp}) \cite{Kresse199607,Kresse199610,Blochl1994,Perdew2009}. Each of the perovskite structure was obtained from a $2\times2\times2$ supercell of the relaxed cubic $AB$O$_3$ perovskite structure. Different phases were obtained by applying appropriate rotation to the $B$O$_6$ octahedra within the supercell. Afterwards, the supercells were biaxially strained on a rigid square grid and the magnitude and angle of the out-of-plane vector, along with the internal coordinates were relaxed. 
The orientation of the biaxially strained structures are shown in Fig. \ref{fig:Strain}. The supercells were rotated such that the (001) ((111)) plane is normal to the $z$ Cartesian coordinate. The source code of \textsc{vasp} was modified to set the stress components on the $x-y$ plane to zero. This allowed for out-of-plane longitudinal and shear stress components to relax the magnitude and direction of the $z$ axis with the $x-y$ plane, along with the internal coordinates.

In order to study the relaxed structures, Crystallographic Information Files (CIF) were generated for each structure using \textsc{findsym} \cite{FINDSYM,Stokes2005}. The CIF files were analysed to obtain irreps of each unstable mode with \textsc{isodistort} \cite{ISODISTORT,Campbell2006}. The order parameters of the Landau Free energy were obtained with \textsc{invariants} \cite{INVARIANTS,Hatch2003}. The Bilbao Crystallographic Server was used as a reference for group theory tables \cite{Aroyo2011}.  Crystal structure visualization was carried out using the \textsc{vesta} software \cite{Momma2011}.

Phonon calculations were performed using the \textit{direct method} as implemented in the \textsc{phonopy} package \cite{Togo2015} on a $3\times3\times3$ supercell. Born charges were obtained from the primitive cell on a $16\times16\times16$ $\Gamma$ centered $k$-point grid.

DFT calculations were performed on a $\Gamma$ centered Monkhorst-Pack grid \cite{Monkhorst1976} of $4\times4\times4$ $k$-points, unless otherwise stated, with a 500 eV energy cutoff for the plane wave basis. Spin-orbit coupling was not included since we found that it did not generate significant qualitative differences. Biaxial strain is measured with respect to the cube root of the computed volume per formula unit of the relaxed $Pnma$ structure, as shown in Table \ref{tab:lc}.

\begin{table}[t]
    \centering
    \begin{tabular}{|c|c|c|}
    \hline
     Compound & \begin{tabular}{c}
     Relaxed structure \\
        volume ({\AA}$^3$) 
     \end{tabular} & Lattice constant ({\AA})\\
     \hline
     CaTiO$_3$ & 220.9 & 3.808\\
     SrSnO$_3$ & 265.5 & 4.049\\
     SrZrO$_3$ & 274.2 & 4.092\\
     \hline
    \end{tabular}
    \caption{Relaxed volumes of the 4 f.u. unitcells of $Pnma$ perovskites, and the corresponding pseudo-cubic lattice parameters calculated by the cube root of a quarter of the cell volumes. These lattice constants are used as the definition of $0\%$ strain. }
    \label{tab:lc}
\end{table}

\section{\label{sec:results}Results and Discussion}

\subsection{\label{sub:instability}Structural Instabilities and Related Phases}

\begin{figure}[ht]
    \centering
    \includegraphics[width=0.99\linewidth]{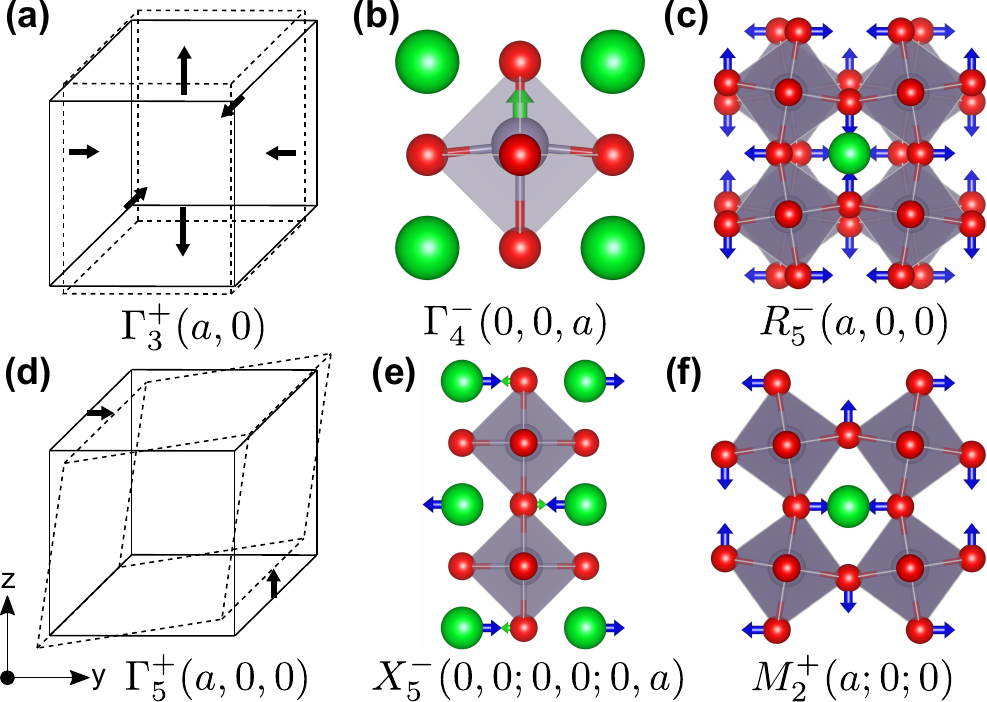}
    \caption{Prominent structural distortions and corresponding irreps present in $Pnma$ perovskites. (a) Normal strain that keep the cell volume constant transforms as $\Gamma_3^+$. (b) Ferroelectric polarization can simply be represented by the displacement of the $B$-site cation, which transforms as $\Gamma_4^-$. (c) $R_5^-$ is the irrep for the out-of-phase octahedral rotation. (d) Shear strain transforms as $\Gamma_5^+$. (e) $X_5^-$ is an antipolar $A$-site displacement and (f) $M_2^+$ is the irrep for the in-phase octahedral rotations.}
    \label{fig:irreps}
\end{figure}

In this section, we give an overview of important types of structural distortions present or can emerge in $Pnma$ perovskites. These structural distortions can be labelled with the irreducible representations (irreps) of the space group, which enables predicting the space groups of various lower symmetry phases, as well as the form of the Landau free energy expansions. The most common instabilities present in the cubic phase of perovskite oxides are the in-phase and out-of-phase octahedral rotations that transform as the $M_2^+$ and $R_5^-$ irreps, shown in Fig.~\ref{fig:irreps}(c) and \ref{fig:irreps}(f). These rotations are are denoted by the `+' and `-' superscripts in the Glazer notation \cite{Glazer1972}, and in perovskites with small tolerance factors (i.e. small $A$-site cations) their combination with the $a^-a^-c^+$ pattern leads to the orthorhombic structure with the $Pnma$ space group. 

The presence of a substrate also breaks the symmetries of the perovskite thin films. In this study, we do not consider interface effects which create a dipole field and directly break the inversion symmetry at the substrate-film interface. We also ignore any `substrate imprinting' effects, which can lower the symmetry of a film further as a result of the structural details of the substrate \cite{Choquette2016}.
In other words, we consider a `strained bulk' configuration, and focus only on strain, which can also be represented by space group irreps. In a cubic material, there are three symmetry inequivalent strain terms: (1) The volume change $\Gamma_1^+$, which is one-dimensional and corresponds to the trace of the strain tensor, (2) the normal strain $\Gamma_3^+$, which is two-dimensional and corresponds to the traceless diagonal part of the strain tensor (Fig.~\ref{fig:irreps}(a)), and (3) the shear strain $\Gamma_5^+$, which is three-dimensional and corresponds to the off-diagonal components of the strain tensor (Fig.~\ref{fig:irreps}(d)). The biaxial strain conditions we consider are nonzero magnitudes of  $\Gamma_3^+$ and $\Gamma_5^+$ for (001) and (111) strains respectively, and they reduce the cubic $Pm\bar{3}m$ symmetry down to tetragonal $P4/mmm$ and rhombohedral $R\bar{3}m$. Table~\ref{tab:strain_group} shows the space groups obtained by the combination of common octahedral rotation patterns with these types of strain.

In addition to the octahedral rotations, perovskites under strain often develop polar instabilities even if they are centrosymmetric (and hence nonpolar) when unstrained. Polarization transforms as the 3D $\Gamma_4^-$ irrep of the cubic reference structure (Fig.~\ref{fig:irreps}(b)). Including the combination of polarization with octahedral rotations and different strain components leads to a large number of possible symmetries listed in Table~\ref{tab:Symmetry}. Additionally, the coexistence of $M_2^+$ and $R_5^-$ rotations can induce many other types of displacements (secondary order parameters) that transform as different irreps. Perhaps the most important one of these displacements is an anti-polar displacement of the $A$-site cations which transform as the $X_5^-$ irrep shown in Fig.~\ref{fig:irreps}(e). This mode is discussed in length especially in the context of hybrid improper ferroelectricity \cite{Mulder2013, Li2021Free}, but it does not lower the symmetry further in perovskites.

\begin{table}[t]
\centering
\begin{tabular}{|c|c|c|c|}
\hline
Rotation    & S.G.      & S.G.              & S.G. \\
Pattern     & (Bulk)    & (001)-strained    & (111)-strained \\
\hline
$a^0a^0a^0$  &  $Pm\bar{3}m$  &  $P4/mmm$  &  $R\bar{3}m$ \\
$a^0a^0c^-$  &  $I4/mcm$  &  $I4/mcm$ &  $C2/c$ \\
$a^-a^-c^0$  &  $Imma$  &  $Imma$ &  $C2/c$ \\
$a^0b^-b^-$  &  $Imma$   &  $C2/m$  &  $C2/c$ \\
$a^-a^-a^-$  &  $R\bar{3}c$  & $C2/c$ &   $R\bar{3}c$  \\
$a^-a^-b^-$  &  $C2/c$  &  $C2/c$  &  $C2/c$  \\
$a^-a^-c^+$  &  $Pnma$   &  $Pnma$  &  $P2_1/c$ \\
$a^+b^-b^-$  &  $Pnma$   &  $P2_1/m$  &  $P2_1/c$ \\
\hline
\end{tabular}
\caption{List of nonpolar space groups considered in this study, obtained by the combinations of octahedral rotations in perovskites, and different strain directions.}
\label{tab:strain_group}
\end{table}

\subsection{\label{sub:001}Phase diagram under (001) strain}

\begin{figure}[ht]
    \centering
    \includegraphics[width=0.8\linewidth]{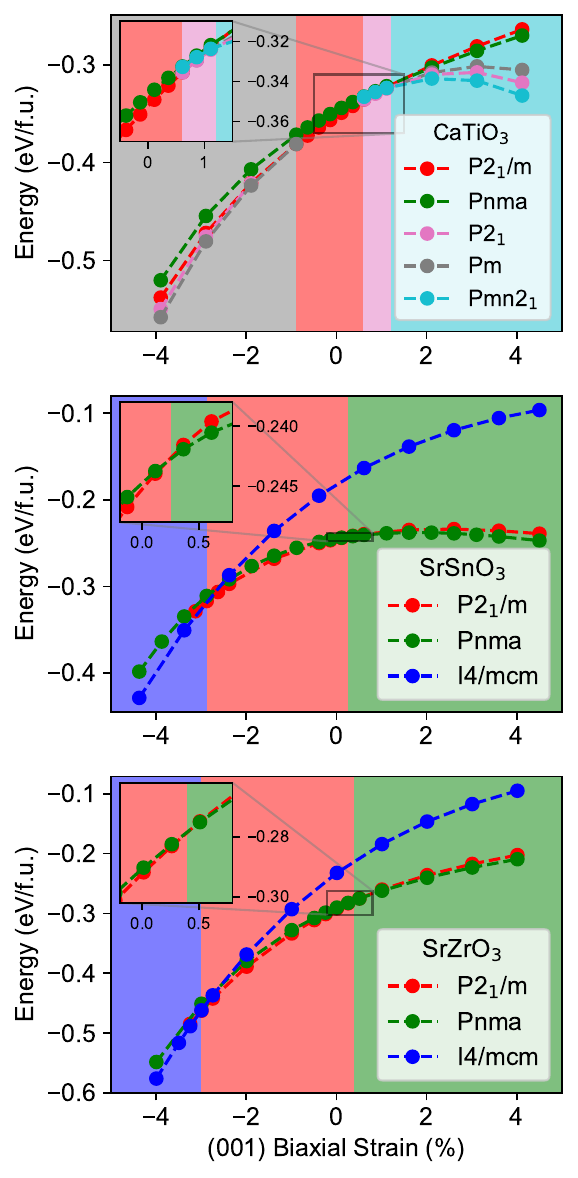}
    \caption{Phase diagram of $Pnma$ Perovskites under (001) epitaxial strain. For each strain value, relative energy stability is reported with respect to the reference $P4/mmm$ structure at the same strain value. Negative (positive) values represent compressive (tensile) (001) biaxial strain. The dotted lines are guides to the eye.}
    \label{fig:Phase001}
\end{figure}

\begin{figure}[ht]
    \centering
    \includegraphics[width=0.98\linewidth]{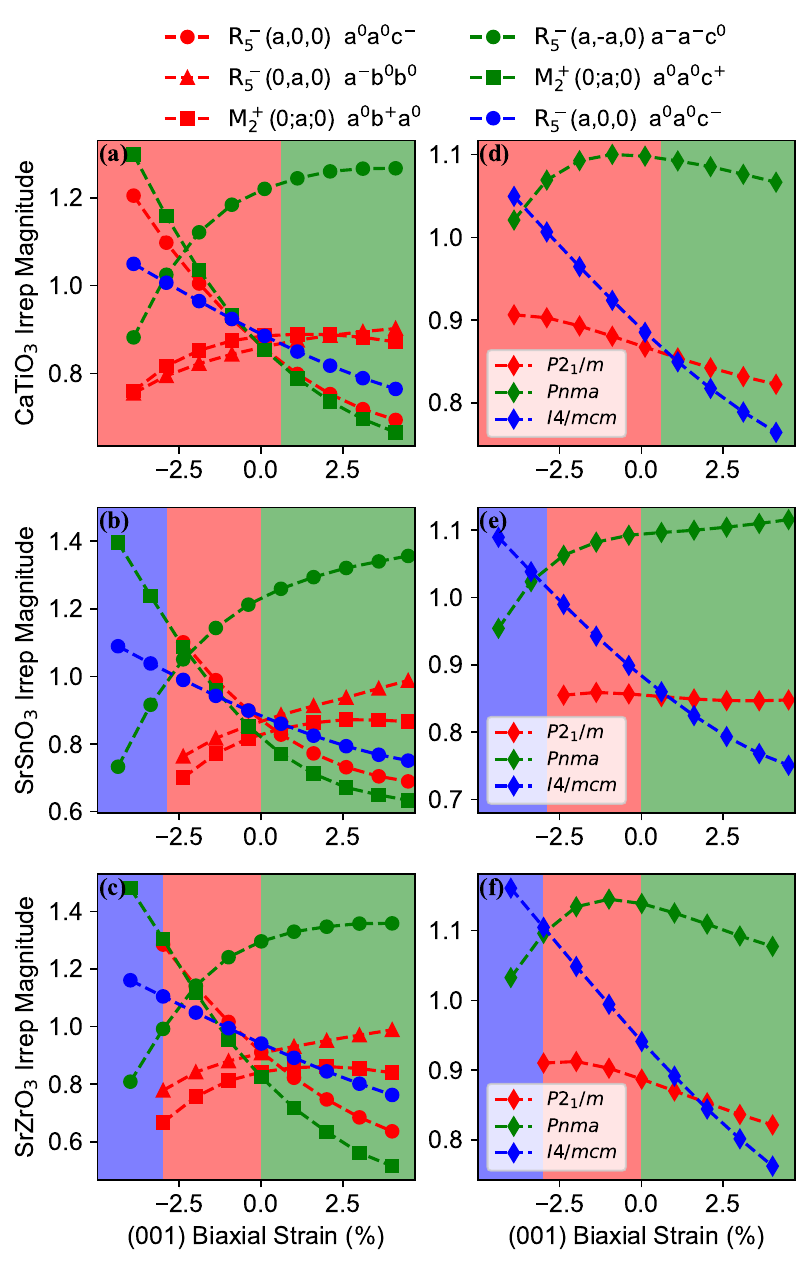}
    \caption{Amplitudes of the octahedral rotation irreps, $R_5^-$ and $M_2^+$, of the stable phases as a function of (001) strain. Panels (a), (b) and (c) show the individual components while panels (d), (e) and (f) show the average rotation magnitude over three axes. The red, green and blue colors correspond to the irreps present in $P2_1/m$, $Pnma$ and $I4/mcm$, respectively. The out-of-plane rotation magnitude increases with increasing (001) compressive strain. This rotation competes with ferroelectric instability, making polarization harder in the compressive region. In the tensile region, the rotation magnitudes do not change significantly.}
    \label{fig:Irrep001}
\end{figure}

Biaxial strain applied in the (001) plane by means of fixing the $a$ and $b$correct pseudocubic lattice constants equal, and the angle between them to 90$^\circ$ transforms as the $\Gamma_3^+(a,0)$ irrep, and reduces the cubic perovskite structure to the tetragonal structure with $P4/mmm$ symmetry. To explore the relative stability of the structures described in Table \ref{tab:strain_group} under (001) strain, we plot the energy gain with respect to the $P4/mmm$ (strained-cubic) structure in Fig. \ref{fig:Phase001}. For simplicity, we only show the ground state candidate structures' energies in this figure, and other phases' energies are presented in the supplementary information \cite{Supplement}.


Not surprisingly, all three compounds display multiple different structural phases as a function of strain. While the energy gain with respect to the $P4/mmm$ structure without any octahedral rotations is always of the order of multiple meV/f.u. (formula unit), there are regions where the two lowest lying states are very close in energy. It is therefore possible that the structural phase is very sensitively dependent on temperature or even quantum fluctuations at zero temperature. Since DFT cannot directly reproduce these fluctuations, we ignore this possibility, as well as a possible phase separation to accommodate strain better, and focus on the lowest energy phases in the remainder of this paper. 

As shown in Fig. \ref{fig:Phase001},  SrSnO$_3$ remains nonpolar within the considered range of strain ($\pm4.5\%$). Under tensile strain, the $Pnma$ structure of SrSnO$_3$ has the lowest energy. This is the same space group as the bulk compound, and it has the axis with the in-phase rotations ($c$) aligned with the normal to the substrate plane. At lower values of tensile, as well as zero and compressive strain values, this is no longer the preferred configuration. Below $\sim0.5\%$ strain, $P2_1/m$ becomes the lower energy structure. In this structure, the rotation pattern is $a^+b^-c^-$, so it can be considered as the same rotation pattern as bulk, but the in-phase rotation axis is aligned in-plane. The reason that this is the lowest energy structure even at zero strain, instead of the $Pnma$ structure, is that the zero strain boundary conditions are different from the bulk zero stress boundary conditions. In particular, the two in-plane axes are fixed to be normal to each other and have the same lattice parameter $a=b$, which is not the case in the orthorhombic $Pnma$ space group in bulk. Also, the zero strain value of $a$ is determined by the cube root of a quarter of the volume of the bulk $Pnma$ structure's unit cell, which is not exactly equal to the pseudo-cubic lattice constant in any direction in the bulk compound. 
At large compressive strain,  around $\sim-3\%$, there is another transition from $P2_1/m$ to $I4/mcm$. This is a single tilt system $a^0a^0c^-$, which is commonly observed in perovskites under compressive strain to accommodate shorter in-plane $B$--O bond-lengths better. 
Consistent with earlier studies \cite{Zhang2017Stannate}, there is no ferroelectric phase in this rather wide strain range. This can be explained by the fact that SrSnO$_3$ has a large $B$-site cation (Sn) that doesn't exhibit significant hybridization between its conduction band minimum (Sn-$s$ orbitals) and the oxygen-$p$ orbitals, unlike most proper ferroelectric transition metal oxides \cite{Bersuker1966, Cohen1990, Benedek2016}. 

Interestingly, the phase diagram of SrZrO$_3$ is qualitatively identical to SrSnO$_3$, despite the fact that Zr is a transition metal ion, and SrZrO$_3$ has a strong ferroelectric instability in its phonon spectrum in the cubic phase \cite{Supplement, Amisi2012} This ferroelectric instability is suppressed by the octahedral rotations in SrZrO$_3$, as is observed in some other $Pnma$ perovskites, and this compound remains nonpolar in the whole strain range we considered ($\pm4\%$). 

CaTiO$_3$ has a very similar phonon spectrum to SrZrO$_3$ in the cubic high-symmetry structure, as discussed in Ref.~\cite{Amisi2012}. It too hosts a polar instability in its parent cubic phase that gets suppressed by the octahedral rotations. However, unlike zirconates, ferroelectricity in titanate perovskites is quite common. BaTiO$_3$ is a prototypical ferroelectric, and both the quantum-paraelectric SrTiO$_3$ and antiferromagnetic EuTiO$_3$ are driven to room temperature ferroelectricity under $\sim1\%$ tensile strain commonly imposed by DyScO$_3$ substrates \cite{Haeni2004, Fennie2006ETO, Lee2010ETO}. Even the layered variants of SrTiO$_3$, the Sr$_{n+1}$Ti$_n$O$_{3n+1}$ Ruddlesden-Poppers have been shown to undergo ferroelectric transitions under strain \cite{Birol2011STO, Lee2013Birol}. 
Molecular beam epitaxy-grown thin films of  CaTiO$_3$ are observed to become ferroelectric under strain \cite{Biegalski2015, Haislmaier2019}, as predicted by DFT \cite{Eklund2009}.

Our DFT-predicted strain phase diagram of (001)-strained CaTiO$_3$ is rife with polar phases, as shown in Fig.~\ref{fig:Phase001}. Around zero strain, biaxially strained CaTiO$_3$ is in nonpolar $P2_1/m$ phase with the $a^+b^-c^-$ rotation pattern. Under about 1\% compressive strain, it undergoes a (second order allowed) phase transition to the polar $Pm$ phase, which arises from polarization of the $P2_1/m$ structure in the out-of-plane direction. Even though this phase is predominantly polarized in the out-of-plane direction, since it is monoclinic, there also exists a small in-plane polarization component. 
In the tensile region, the $P2_1/m$ phase undergoes another second order-allowed phase transition to the polar $P2_1$ phase, which has an in-plane polarization of the $P2_1/m$ structure, at around 0.5\% strain. At around 1\% tensile strain, there is a first order phase transition to $Pmn2_1$ phase, which arises from in-plane polarization of $Pnma$ structure. Interestingly, we find that the space group of the bulk compound, $Pnma$ is not realized at any strain value in thin films because of the different boundary conditions for CaTiO$_3$.

A pattern that is common in these three $Pnma$ perovskites is that under tensile strain, they have the $a^-a^-c^+$ octahedral rotation pattern, and hence the $Pnma$ or its in-plane polar subgroup in the tensile region; whereas they have the $a^+b^-c^-$ rotation pattern and and either $P2_1/m$ or its out-of-plane polar subgroup under compressive (001) strain. 
In Fig.~\ref{fig:Irrep001}, we show the magnitudes of different components of the octahedral rotations decomposed into the $R_5^-$ and $M_2^+$ irreps. A trend that is evident is that the rotations around the out-of-plane axis increase rapidly under increasing compressive strain in both the $Pnma$ or the $I4/mcm$ space group. This is not a surprising observation, since this sign of strain decreases the $B$--O bond-length, which can be increased back by the larger octahedral rotation angle that it holds under $Pnma$ (and even the metastable $P2_1/m$) structure as well. 

\begin{figure}[t]
    \centering
    \includegraphics[width=0.8\linewidth]{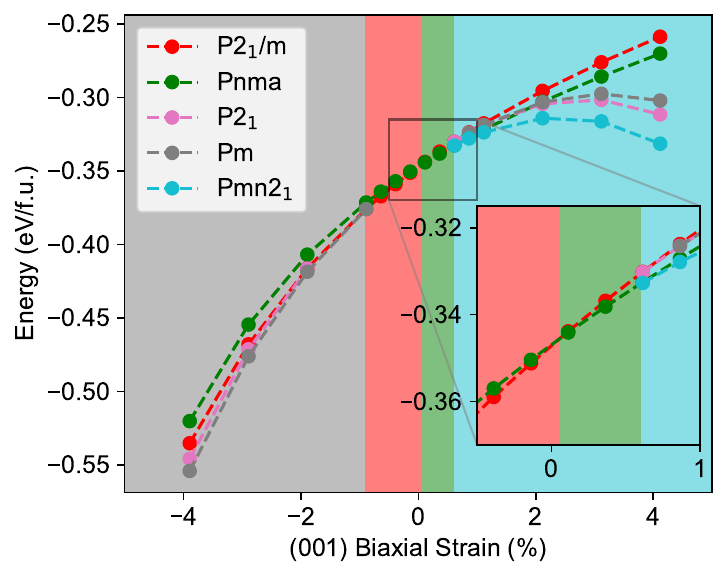}
    \caption{Phase diagram of CaTiO$_3$ under (001) biaxial strain without relaxing the monoclinic angles the $c$ axis makes with the biaxial grid.}
    \label{fig:compare}
\end{figure}

In passing, we underline the importance of relaxing the monoclinic angle that the out-of-plane lattice vectors make with the in-plane ones in biaxial strain calculations. Especially for small strain values, this angle does not differ much from $90^\circ$, and is often assumed to be constant for simplicity. However, especially in regions where multiple phases are close in energy, this angle can make a difference. In Fig.~\ref{fig:compare}, we show the phase diagram of CaTiO$_3$ without the monoclinic angle relaxed. When the monoclinic angle is not relaxed, the orthorhombic $Pnma$ phase is favored, and it is stabilized in a narrow strain range under small tensile strain, as opposed to the results in Fig.~\ref{fig:Phase001}(c). This observation signifies the importance of relaxing the monoclinic angle to obtain the correct ground state.

\subsection{\label{sub:111}Phase diagram under (111) strain}

In this section, we consider (111) films. We impose the (111) strain in a way similar to the (001) strain by building a $2\times2\times2$ supercell of the cubic perovskite structure, changing and fixing the lattice vectors' components on the (111) plane, and relaxing all the other degrees of freedom after applying various octahedral rotation to the $B$O$_6$ unit. The strain on the  (111) plane corresponds to a shear and transforms as the $\Gamma_5^+$ irrep, reducing the symmetry to rhombohedral $R\bar{3}m$. As a result, we use the $R\bar{3}m$ structure as reference and obtain the energies of other structures, with octahedral rotations, with respect the reference $R\bar{3}m$ structure of equal strain. 

\begin{figure}[t]
    \centering
    \includegraphics[width=0.8\linewidth]{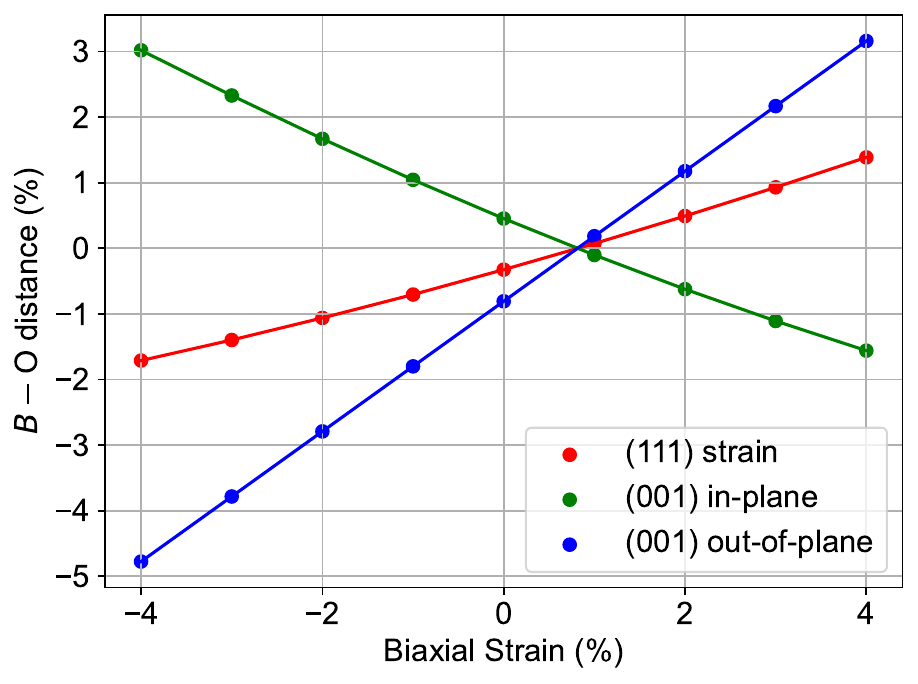}
    \caption{The $B-$O distance as a function of biaxial strain for cubic CaTiO$_3$. The red, green and blue lines correspond to the Ti--O distances under (111) strain (rhombohedral $R\bar{3}m$) and (001) strain (orthorhombic $P4/mmm$ with two inequivalent Ti--O distances). The lines meet at a point where the structure is cubic and changes in Ti--O distance is in reference to the this structure.}
    \label{fig:distance}
\end{figure}

\begin{figure}[ht]
    \centering
    \includegraphics[width=0.8\linewidth]{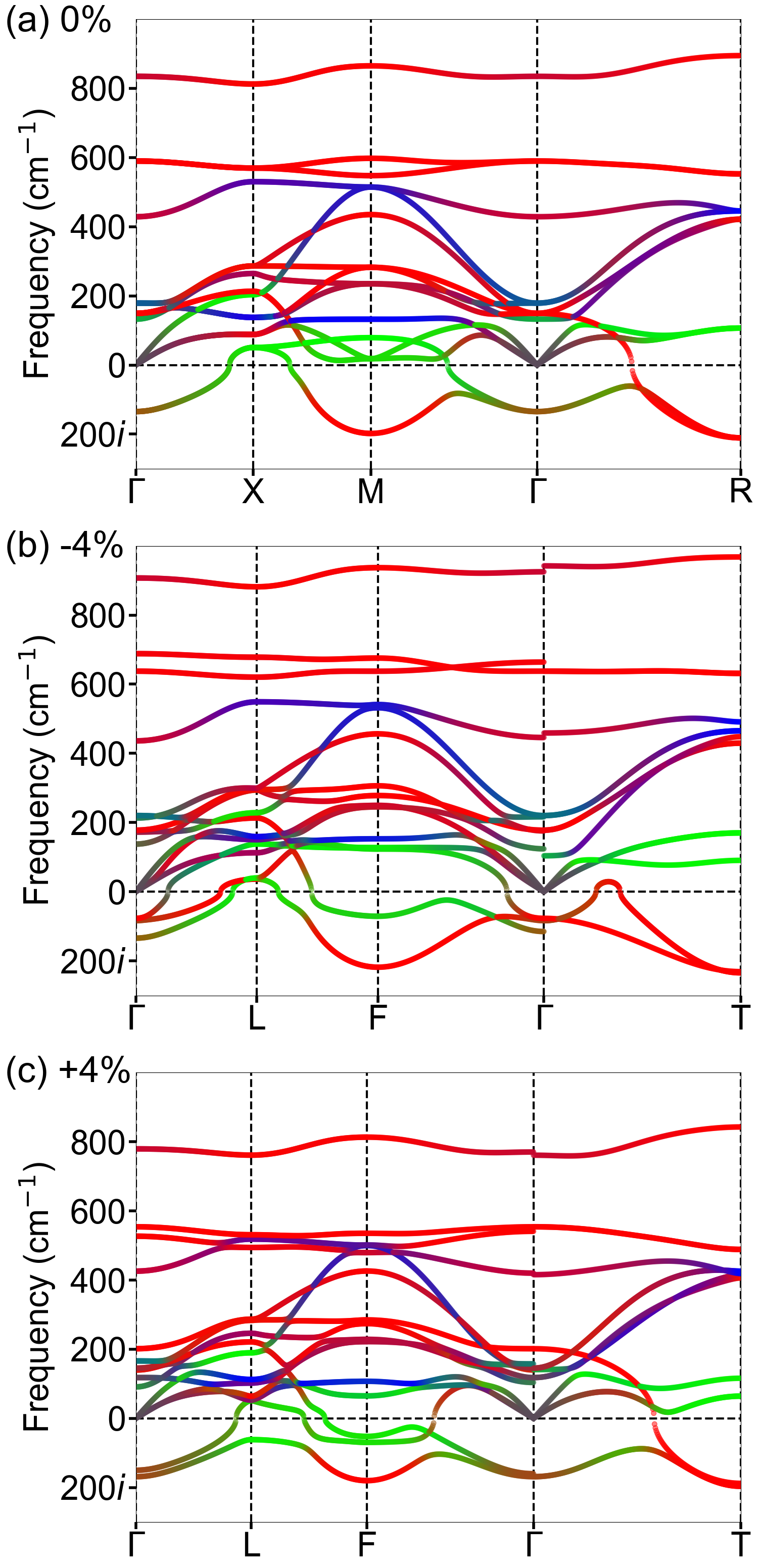}
    \caption{Phonon dispersions of CaTiO$_3$ for (a) unstrained cubic structure, and (111)-strained structures with $R\bar{3}m$ symmetry at (b) $-4\%$ and (c) $+4\%$ strain. The colors denote the atomic displacement magnitude for Ca (green), Ti (blue) and O (red). We compare the phonon dispersion along related $k$-points $\Gamma = (0,0,0)$, $X = (0.5,0,0) = L$, $M = (0.5,0.5,0) = F$ and $R = (0.5,0.5,0.5) =T$ between the cubic and rhombohedral structures.}
    \label{fig:phonon}
\end{figure}

Note that while a single lattice parameter (the length of the $c$-axis) is relaxed on a common (001) strain calculation without any structural distortions besides strain, the (111) strain calculation relax the length of the lattice vector along the [111], which can involve both a change of the rhombohedral angle or the lattice parameter in the rhombohedral structure. 
In other words, for a cubic perovskite without structural distortions, the (111) strain boundary conditions leave two degrees of freedom as free parameters, as opposed to a single free parameter under the (001) strain boundary conditions. This makes it easier for the compound to accommodate bond lengths without reducing symmetry. This point is illustrated in Fig.~\ref{fig:distance}, which shows the $Ti-O$ bond length in CaTiO$_3$ with no octahedral rotations or polarization under strain. The relaxed structure under different values of (111) strain has its $Ti-O$ bond lengths remain equal to each other and depend on strain at a rate of $\lesssim0.4$ (red line). The (001) strain, on the other hand, leads to two unequal in-plane and out-of-plane bond lengths, which change at rates of $\sim -0.6$ and $\sim 1.0$ respectively (green and blue lines). 
As a result of this degree of freedom, it is likely harder to create new strong instabilities via (111) strain compared to (001) strain. The phonon dispersions of CaTiO$_3$, shown in Fig.~\ref{fig:phonon} for $0\%$ and $\mp4\%$ strain, supports this expectation: While there are some changes in the phonon spectrum, even this large amount of strain doesn't make a dramatic difference in the nature or frequency of the strongest instabilities, which remain the octahedral rotational instabilities.

\begin{figure}[ht]
    \centering
    \includegraphics[width=0.8\linewidth]{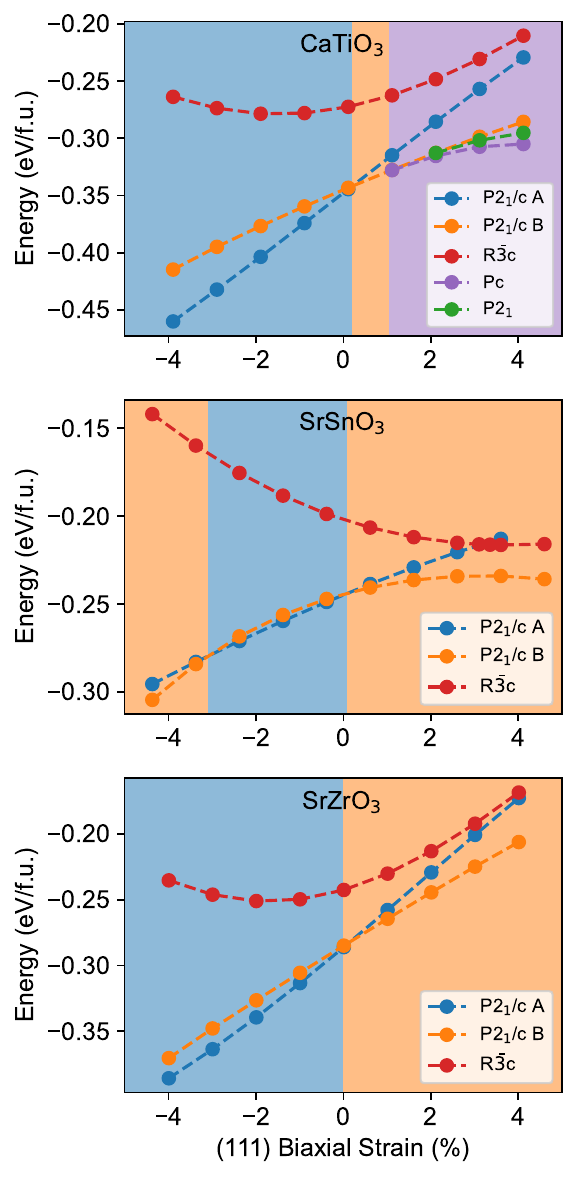}
    \caption{Phase diagram of $Pnma$ Perovskites under (111) epitaxial strain. For each strain value, relative energy stability is computed with respect to the reference R$\bar{3}$m structure at the same strain value. Negative (positive) values represent compressive (tensile) (111) biaxial strain. The dotted lines are guides to the eye. The phase diagram is dominated by two distinct structures, both having the $P2_1/c$ symmetry.
    }
    \label{fig:Phase111}
\end{figure}

\begin{figure}[ht]
    \centering
    \includegraphics[width=0.98\linewidth]{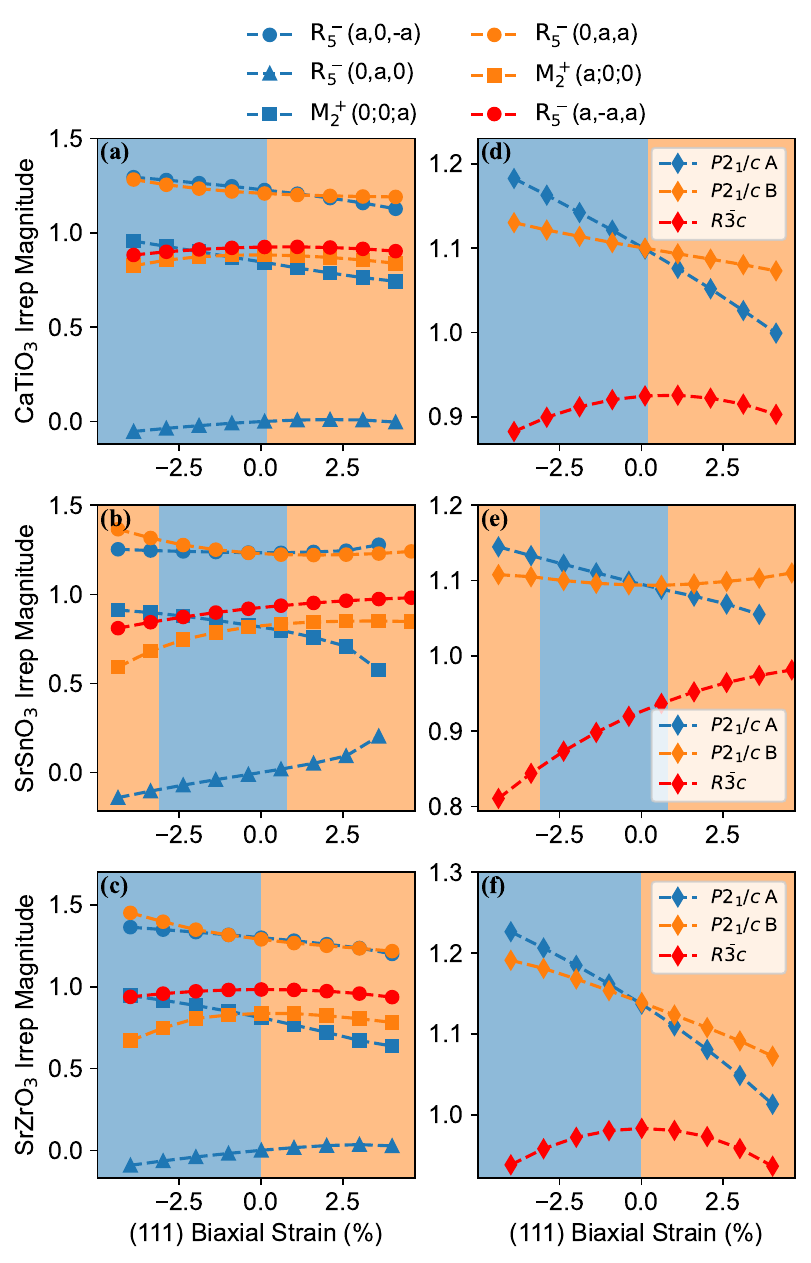}
    \caption{Octahedral rotation irreps, $R_5^-$ and $M_2^+$, of the relevant phases as a function of (111) strain. Panels (a), (b) and (c) show the individual components while panels (d), (e) and (f) show the average rotation magnitude over three axes. The blue, orange and red colors correspond to the irreps of $P2_1/c$ Phase A, $2_1/c$ Phase B and $R\bar{3}c$, respectively. The magnitude of the rotation gradually increases when going from tensile to compressive (111) strain. The phase with larger rotation magnitude is generally more stable since it provides better stabilization to the small $A$-site.}
    \label{fig:Irrep111}
\end{figure}

The phase diagram under (111) strain is shown in Fig. \ref{fig:Phase111}. SrZrO$_3$ realize two different phases under opposite signs of strain, both with $P2_1/c$ symmetry. (Note that this symmetry is obtained trivially when $Pnma$ perovskites are placed under a (111) strain as shown in Tables~\ref{tab:strain_group} and \ref{tab:Symmetry}.) We find that although these two states have the same space group name, they are distinct structures with different directions for the primary $M_2^+$ and $R_5^-$ order parameters, as well as different induced secondary order parameter irreps. They also have different origin and bases with respect to the parent $Pm\bar{3}m$ structure. Hence, the transition at $\sim0\%$ strain is not an isostructural transition, despite the \textit{name} of the spacegroup remaining the same. We denote the two phases with spacegroup $P2_1/c$ as Phase A in the compressive region and Phase B in the tensile region. In Fig.~\ref{fig:Irrep111} and the supplementary information \cite{Supplement}, we show a decomposition of both phases into the structural distortions they host. While there is a complex list of many distortions with different irreps in both structures, a distinguishing feature, for example, is the presence of small amplitude secondary order parameters $R_2^-$ and $M_2^-$ in Phase B. 

SrSnO$_3$ also has only the same two phases and a phase transition between them near $\sim0\%$ strain. However, interestingly, the ground state of SrSnO$_3$ becomes Phase-B again at large compressive strain. There is no obvious explanation of this behavior. The transition near $0\%$ in both compounds seem to arise from a preference for a phase with larger rotation angles (most clearly seen on the right panels of Fig.~\ref{fig:Irrep111}), but even the complete irrep decomposition of SrSnO$_3$ (left panels of Fig.~\ref{fig:Irrep111}, as well as Supplementary Fig.~20, 22) doesn't point to any clues about the energy trends at larger compressive strain.

CaTiO$_3$ also has the same phase transition around 0\% strain, and so we conclude that the presence of two different $Pnma$-like $P2_1/c$ phases is a general property of many $Pnma$ perovskites under (111) strain. Additionally, in-line with the stronger ferroelectric tendencies of CaTiO$_3$ discussed earlier, there is another, second-order allowed phase transition at $\sim+1\%$ tensile strain from $P2_1/c$ Phase B to the predominantly-in-plane polar $Pc$ phase. This phase is again distinct from the polar $Pc$ phase (not shown) obtained from the Phase A of $P2_1/c$ even though its space group name is the same.

\begin{table*}[t]
    \centering
    \begin{tabular}{|l c c| c c |c c|}
    \hline
    \hline
     \begin{tabular}{c}
     Bulk  \\
     phase 
     \end{tabular} &  
     \begin{tabular}{c}
     Glazer  \\
     notation 
     \end{tabular} & 
     \begin{tabular}{c}
     Symm.  \\
     modes 
     \end{tabular} & 
     \begin{tabular}{c}
     (111)  \\
     strain 
     \end{tabular} & Polar phases & 
     \begin{tabular}{c}
     (001)  \\
     strain 
     \end{tabular} & Polar phases \\
    \hline
     $Pm\bar{3}m$  &  a$^0$a$^0$a$^0$  &  &  $R\bar{3}m$  &  $R3m$; $C2$; $Cm$  &  $P4/mmm$  &  $Pmm2$; $P4mm$; $Amm2$; $Pm$; $Cm$ \\
     $I4/mcm$  &  a$^0$a$^0$b$^-$  &  $R_5^-$  &  $C2/c$  &  $C2$; $Cc$  &  $I4/mcm$  &  $Fmm2$; $I4cm$; $Ima2$; $Cm$; $Cc$ \\
     &  a$^-$b$^0$b$^0$  &  &  &  &  $Fmmm$  &  $Fmm2$; $Cm$ \\
     $Imma$  &  a$^-$a$^-$b$^0$  &  $R_5^-$  &  $C2/c$  &  $C2$; $Cc$  &  $Imma$  &  $Cm$; $Ima2$; $Imm2$; $Cc$ \\
     &  a$^0$b$^-$b$^-$  &  &  &  &  $C2/m$ &  $C2$; $Cm$ \\
     $R\bar{3}c$  &  a$^-$a$^-$a$^-$  &  $R_5^-$  &  $R\bar{3}c$  &  $R3c$; $C2$; $Cc$  &  $C2/c$  &   $C2$; $Cc$ \\
     $P4/mbm$  &  a$^0$a$^0$b$^+$  &  $M_2^+$  &  $P2_1/c$  &  $P2_1$; $Pc$  &  $P4/mbm$  &  $Amm2$; $P4bm$; $Pmc2_1$; $Pm$; $Cm$; $Pc$ \\
     &  a$^+$b$^0$b$^0$  &  &  &  &  $Cmmm$  &  $Cmm2$; $Amm2$; $Cm$; $Pm$ \\
     $I4/mmm$  &  a$^+$a$^+$b$^0$  &  $M_2^+$  &  $C2/m$  &  $C2$; $Cm$  &  $I4/mmm$  &  $Imm2$; $I4mm$; $Fmm2$; $Cm$  \\
     &  a$^0$b$^+$b$^+$  &  &  &  &  $Immm$  &  $Imm2$; $Cm$ \\
     $Im\bar{3}$  &  a$^+$a$^+$a$^+$  &  $M_2^+$  &  $R\bar{3}$  &  $R3$  &  $Immm$  &  $Imm2$; $Cm$ \\
     $C2/m$  &  a$^0$b$^-$c$^-$  &  $R_5^-$  &  $P\bar{1}$  &  &  $C2/m$  &  $C2$; $Cm$ \\
     $C2/c$  &  a$^-$a$^-$b$^-$  &  $R_5^-$  &  $C2/c$  &  $C2$; $Cc$  &  $C2/c$  &  $C2$; $Cc$  \\
     &  a$^-$b$^-$b$^-$  &  &  &  &  $P\bar{1}$  &  \\
     $Pnma$   &  a$^-$a$^-$b$^+$  &  $M_2^+$,$R_5^-$  &  $P2_1/c$  &  $P2_1$; $Pc$  &  $Pnma$   &  $Pna2_1$; $Pmc2_1$; $Pmn2_1$; $Pm$; $Pc$ \\
     &  a$^+$b$^-$b$^-$  &  &  &  &  $P2_1/m$  &  $P2_1$; $Pm$ \\
     $Cmcm$  &  a$^0$b$^+$c$^-$  &  $M_2^+$,$R_5^-$  &  $P\bar{1}$  &  &  $Cmcm$  &  $Ama2$; $Cmc2_1$; $Amm2$; $Cc$; $Pm$; $Cm$ \\
     $P4_2/nmc$  &  a$^+$a$^+$b$^-$  &  $M_2^+$,$R_5^-$  &  $C2/c$  &  $C2$; $Cc$  &  $P4_2/nmc$  &  $Pmn2_1$; $P4_2mc$; $Aba2$; $Pc$; $Pm$; $Cc$ \\
     &  a$^-$b$^+$b$^+$  &  &  &  &  $Pmnm$  &  $Pmm2$; $Pmn2_1$; $Pm$; $Pc$ \\
     $Immm$  &  a$^+$b$^+$c$^+$  &  $M_2^+$  &  $P\bar{1}$  &  &  $Immm$  &  $Imm2$; $Cm$ \\
     $P2_1/m$  &  a$^+$b$^-$c$^-$  &  $M_2^+$,$R_5^-$  &  $P\bar{1}$  &  &  $P2_1/m$  &  $P2_1$; $Pm$ \\
     $P\bar{1}$  &  a$^-$b$^-$c$^-$  &  $R_5^-$  &  $P\bar{1}$  &  &  $P\bar{1}$  &  \\
     \hline
    \end{tabular}
    \caption{Symmetries of all perovskite phases with different rotation patterns, polarization, and biaxial strain.  We list the rotation patterns in Glazer notation, relevant rotation irreps, strained phases and their polar phases obtained under (001) strain ($\Gamma_3^+$), (111) strain ($\Gamma_5^+$) and polarization along cubic axes ($\Gamma_4^-$). Symmetry modes are given with respect to the cubic $Pm\bar{3}m$ structure of $AB$O$_3$ perovskite with atomic position given by $A$ (1a), $B$ (1b), and O (3c). The polar structure $P1$ with polarization along a general direction is trivially allowed in each row and hence is not listed.}
    \label{tab:Symmetry}
\end{table*}

\section{\label{sub:discussion}Discussion}

There are distinct differences between the phase diagrams of $Pnma$ perovskites under (001) and (111) epitaxial strain. Firstly, there is a larger number of different phases present in the (001) case. This is in part because of the greater richness of the phase space in the (001) case as seen from Table~\ref{tab:Symmetry}, which lists the symmetries of all possible structures under (001) and (111) strain. Due to the fact that the axes of the octahedral rotation in $Pnma$ are aligned with a unique crystalline axis (the [001] direction), it can achieve a larger pool of structures with different space groups, by permuting the direction of rotation and strain. The (111) strain applies for all three possible pseudo-cubic directions for octahedral rotations equally, and hence, can only achieve one symmetry configuration for each octahedral rotation pattern.
Another reason that helps having fewer distinct phases under (111) strain is the seemingly lower sensitivity of the phonon instabilities to (111) biaxial strain due to strain itself not affecting the lattice vectors as much as (001) strain. 

Secondly, CaTiO$_3$ polarizes readily under (001) strain, in both the compressive and tensile region. Under (111), it only polarizes in the tensile region, and only at larger strain magnitude. A similar situation was observed from DFT in SrTiO$_3$ \cite{Reyes-Lillo2019} which has weak $a^0a^0c^-$ rotations and becomes polar under tensile (111) strain only, and BaTiO$_3$ \cite{Oja2008} which is polar in bulk but becomes centrosymmetric under compressive (111) strain.

\section{\label{sec:conclusion}Summary and Conclusions}

In this study, we applied first principles DFT to investigate the effect of biaxial strain on (001) and (111)-oriented $Pnma$ perovskites, which constitute more than half of all perovskite oxides \cite{Lufaso2001}. We found that it is essential to relax the monoclinic angles of the out-of-plane vectors while applying the biaxial constraint, especially when monoclinic phases are abundant in the phase diagram, since it plays an important role in stabilizing these phases and hence obtaining the correct ground state structure. Under (111) strain, we found two different structures in the tensile and compressive strain regions, both having $P2_1/c$ symmetry. We also found that CaTiO$_3$ polarizes under small tensile and compressive (001) strain and relatively larger tensile (111) strain. The phase diagram for CaTiO$_3$ under (001) strain is differs from previous predictions \cite{Eklund2009}, with no $Pnma$ phase and polarization in the compressive strain region. 

We performed an analysis of all distortions with different irreps present in these structures as a function of strain and found that the magnitude of irreps remain relatively stable under (111) strain. For (001) strain, many octahedral rotational modes' amplitudes reduce with increasing tensile strain, and increase with a relatively high rate under compressive strain.

In summary, we compared the effects of two distinct strain directions in biaxial perovskites, and showed that the (111) strain not only realize symmetries that are not present in (001) strained films, but also the trends of octahedral rotation angles and ferroelectric polarization is very different under (111) strain than it is under (001) strain. Hence, different strain directions is a way to manipulate the crystal structure and obtain new phases of complex oxides. While all compounds we considered are band insulators, these new structural phases' effects on the electronic structure will likely provide many open questions of increasing importance due to the increasing availability of (111) films grown by molecular beam epitaxy or pulsed laser deposition.

\acknowledgements

This work was supported by the US Department of Energy through the University of Minnesota (UMN) Center for Quantum Materials, under Grant No. DE-SC0016371.


\begin{thebibliography}{66}%
\makeatletter
\providecommand \@ifxundefined [1]{%
 \@ifx{#1\undefined}
}%
\providecommand \@ifnum [1]{%
 \ifnum #1\expandafter \@firstoftwo
 \else \expandafter \@secondoftwo
 \fi
}%
\providecommand \@ifx [1]{%
 \ifx #1\expandafter \@firstoftwo
 \else \expandafter \@secondoftwo
 \fi
}%
\providecommand \natexlab [1]{#1}%
\providecommand \enquote  [1]{``#1''}%
\providecommand \bibnamefont  [1]{#1}%
\providecommand \bibfnamefont [1]{#1}%
\providecommand \citenamefont [1]{#1}%
\providecommand \href@noop [0]{\@secondoftwo}%
\providecommand \href [0]{\begingroup \@sanitize@url \@href}%
\providecommand \@href[1]{\@@startlink{#1}\@@href}%
\providecommand \@@href[1]{\endgroup#1\@@endlink}%
\providecommand \@sanitize@url [0]{\catcode `\\12\catcode `\$12\catcode
  `\&12\catcode `\#12\catcode `\^12\catcode `\_12\catcode `\%12\relax}%
\providecommand \@@startlink[1]{}%
\providecommand \@@endlink[0]{}%
\providecommand \url  [0]{\begingroup\@sanitize@url \@url }%
\providecommand \@url [1]{\endgroup\@href {#1}{\urlprefix }}%
\providecommand \urlprefix  [0]{URL }%
\providecommand \Eprint [0]{\href }%
\providecommand \doibase [0]{https://doi.org/}%
\providecommand \selectlanguage [0]{\@gobble}%
\providecommand \bibinfo  [0]{\@secondoftwo}%
\providecommand \bibfield  [0]{\@secondoftwo}%
\providecommand \translation [1]{[#1]}%
\providecommand \BibitemOpen [0]{}%
\providecommand \bibitemStop [0]{}%
\providecommand \bibitemNoStop [0]{.\EOS\space}%
\providecommand \EOS [0]{\spacefactor3000\relax}%
\providecommand \BibitemShut  [1]{\csname bibitem#1\endcsname}%
\let\auto@bib@innerbib\@empty
\bibitem [{\citenamefont {Moloney}\ \emph {et~al.}(2020)\citenamefont
  {Moloney}, \citenamefont {Yeddu},\ and\ \citenamefont
  {Saidaminov}}]{Moloney2020}%
  \BibitemOpen
  \bibfield  {author} {\bibinfo {author} {\bibfnamefont {E.~G.}\ \bibnamefont
  {Moloney}}, \bibinfo {author} {\bibfnamefont {V.}~\bibnamefont {Yeddu}},\
  and\ \bibinfo {author} {\bibfnamefont {M.~I.}\ \bibnamefont {Saidaminov}},\
  }\bibfield  {title} {\bibinfo {title} {Strain engineering in halide
  perovskites},\ }\href@noop {} {\bibfield  {journal} {\bibinfo  {journal} {ACS
  Materials Letters}\ }\textbf {\bibinfo {volume} {2}},\ \bibinfo {pages}
  {1495} (\bibinfo {year} {2020})}\BibitemShut {NoStop}%
\bibitem [{\citenamefont {Schlom}\ \emph {et~al.}(2014)\citenamefont {Schlom},
  \citenamefont {Chen}, \citenamefont {Fennie}, \citenamefont {Gopalan},
  \citenamefont {Muller}, \citenamefont {Pan}, \citenamefont {Ramesh},\ and\
  \citenamefont {Uecker}}]{Schlom2014}%
  \BibitemOpen
  \bibfield  {author} {\bibinfo {author} {\bibfnamefont {D.~G.}\ \bibnamefont
  {Schlom}}, \bibinfo {author} {\bibfnamefont {L.-Q.}\ \bibnamefont {Chen}},
  \bibinfo {author} {\bibfnamefont {C.~J.}\ \bibnamefont {Fennie}}, \bibinfo
  {author} {\bibfnamefont {V.}~\bibnamefont {Gopalan}}, \bibinfo {author}
  {\bibfnamefont {D.~A.}\ \bibnamefont {Muller}}, \bibinfo {author}
  {\bibfnamefont {X.}~\bibnamefont {Pan}}, \bibinfo {author} {\bibfnamefont
  {R.}~\bibnamefont {Ramesh}},\ and\ \bibinfo {author} {\bibfnamefont
  {R.}~\bibnamefont {Uecker}},\ }\bibfield  {title} {\bibinfo {title} {Elastic
  strain engineering of ferroic oxides},\ }\href
  {https://doi.org/10.1557/mrs.2014.1} {\bibfield  {journal} {\bibinfo
  {journal} {MRS Bulletin}\ }\textbf {\bibinfo {volume} {39}},\ \bibinfo
  {pages} {118–130} (\bibinfo {year} {2014})}\BibitemShut {NoStop}%
\bibitem [{\citenamefont {Zhu}\ \emph {et~al.}(2019)\citenamefont {Zhu},
  \citenamefont {Niu}, \citenamefont {Fu}, \citenamefont {Li}, \citenamefont
  {Hu}, \citenamefont {Chen}, \citenamefont {He}, \citenamefont {Na},
  \citenamefont {Liu}, \citenamefont {Zai} \emph {et~al.}}]{Zhu2019}%
  \BibitemOpen
  \bibfield  {author} {\bibinfo {author} {\bibfnamefont {C.}~\bibnamefont
  {Zhu}}, \bibinfo {author} {\bibfnamefont {X.}~\bibnamefont {Niu}}, \bibinfo
  {author} {\bibfnamefont {Y.}~\bibnamefont {Fu}}, \bibinfo {author}
  {\bibfnamefont {N.}~\bibnamefont {Li}}, \bibinfo {author} {\bibfnamefont
  {C.}~\bibnamefont {Hu}}, \bibinfo {author} {\bibfnamefont {Y.}~\bibnamefont
  {Chen}}, \bibinfo {author} {\bibfnamefont {X.}~\bibnamefont {He}}, \bibinfo
  {author} {\bibfnamefont {G.}~\bibnamefont {Na}}, \bibinfo {author}
  {\bibfnamefont {P.}~\bibnamefont {Liu}}, \bibinfo {author} {\bibfnamefont
  {H.}~\bibnamefont {Zai}}, \emph {et~al.},\ }\bibfield  {title} {\bibinfo
  {title} {Strain engineering in perovskite solar cells and its impacts on
  carrier dynamics},\ }\href@noop {} {\bibfield  {journal} {\bibinfo  {journal}
  {Nature communications}\ }\textbf {\bibinfo {volume} {10}},\ \bibinfo {pages}
  {815} (\bibinfo {year} {2019})}\BibitemShut {NoStop}%
\bibitem [{\citenamefont {Jiao}\ \emph {et~al.}(2021)\citenamefont {Jiao},
  \citenamefont {Yi}, \citenamefont {Wang}, \citenamefont {Li}, \citenamefont
  {Hao}, \citenamefont {Pan}, \citenamefont {Shi}, \citenamefont {Li},
  \citenamefont {Liu}, \citenamefont {Zhang} \emph {et~al.}}]{Jiao2021}%
  \BibitemOpen
  \bibfield  {author} {\bibinfo {author} {\bibfnamefont {Y.}~\bibnamefont
  {Jiao}}, \bibinfo {author} {\bibfnamefont {S.}~\bibnamefont {Yi}}, \bibinfo
  {author} {\bibfnamefont {H.}~\bibnamefont {Wang}}, \bibinfo {author}
  {\bibfnamefont {B.}~\bibnamefont {Li}}, \bibinfo {author} {\bibfnamefont
  {W.}~\bibnamefont {Hao}}, \bibinfo {author} {\bibfnamefont {L.}~\bibnamefont
  {Pan}}, \bibinfo {author} {\bibfnamefont {Y.}~\bibnamefont {Shi}}, \bibinfo
  {author} {\bibfnamefont {X.}~\bibnamefont {Li}}, \bibinfo {author}
  {\bibfnamefont {P.}~\bibnamefont {Liu}}, \bibinfo {author} {\bibfnamefont
  {H.}~\bibnamefont {Zhang}}, \emph {et~al.},\ }\bibfield  {title} {\bibinfo
  {title} {Strain engineering of metal halide perovskites on coupling
  anisotropic behaviors},\ }\href@noop {} {\bibfield  {journal} {\bibinfo
  {journal} {Advanced Functional Materials}\ }\textbf {\bibinfo {volume}
  {31}},\ \bibinfo {pages} {2006243} (\bibinfo {year} {2021})}\BibitemShut
  {NoStop}%
\bibitem [{\citenamefont {Liu}\ \emph {et~al.}(2022)\citenamefont {Liu},
  \citenamefont {Li}, \citenamefont {Wang}, \citenamefont {Ye}, \citenamefont
  {Yan}, \citenamefont {Zhang}, \citenamefont {Dong}, \citenamefont {Lu},
  \citenamefont {Huang}, \citenamefont {He} \emph {et~al.}}]{Liu2022}%
  \BibitemOpen
  \bibfield  {author} {\bibinfo {author} {\bibfnamefont {B.}~\bibnamefont
  {Liu}}, \bibinfo {author} {\bibfnamefont {J.}~\bibnamefont {Li}}, \bibinfo
  {author} {\bibfnamefont {G.}~\bibnamefont {Wang}}, \bibinfo {author}
  {\bibfnamefont {F.}~\bibnamefont {Ye}}, \bibinfo {author} {\bibfnamefont
  {H.}~\bibnamefont {Yan}}, \bibinfo {author} {\bibfnamefont {M.}~\bibnamefont
  {Zhang}}, \bibinfo {author} {\bibfnamefont {S.-C.}\ \bibnamefont {Dong}},
  \bibinfo {author} {\bibfnamefont {L.}~\bibnamefont {Lu}}, \bibinfo {author}
  {\bibfnamefont {P.}~\bibnamefont {Huang}}, \bibinfo {author} {\bibfnamefont
  {T.}~\bibnamefont {He}}, \emph {et~al.},\ }\bibfield  {title} {\bibinfo
  {title} {Lattice strain modulation toward efficient blue perovskite
  light-emitting diodes},\ }\href@noop {} {\bibfield  {journal} {\bibinfo
  {journal} {Science Advances}\ }\textbf {\bibinfo {volume} {8}},\ \bibinfo
  {pages} {eabq0138} (\bibinfo {year} {2022})}\BibitemShut {NoStop}%
\bibitem [{\citenamefont {Yang}\ and\ \citenamefont {Huo}(2020)}]{Yang2020}%
  \BibitemOpen
  \bibfield  {author} {\bibinfo {author} {\bibfnamefont {D.}~\bibnamefont
  {Yang}}\ and\ \bibinfo {author} {\bibfnamefont {D.}~\bibnamefont {Huo}},\
  }\bibfield  {title} {\bibinfo {title} {Cation doping and strain engineering
  of {CsPbBr$_3$}-based perovskite light emitting diodes},\ }\href@noop {}
  {\bibfield  {journal} {\bibinfo  {journal} {Journal of Materials Chemistry
  C}\ }\textbf {\bibinfo {volume} {8}},\ \bibinfo {pages} {6640} (\bibinfo
  {year} {2020})}\BibitemShut {NoStop}%
\bibitem [{\citenamefont {Dimos}\ and\ \citenamefont
  {Mueller}(1998)}]{Dimos1998}%
  \BibitemOpen
  \bibfield  {author} {\bibinfo {author} {\bibfnamefont {D.}~\bibnamefont
  {Dimos}}\ and\ \bibinfo {author} {\bibfnamefont {C.}~\bibnamefont
  {Mueller}},\ }\bibfield  {title} {\bibinfo {title} {Perovskite thin films for
  high-frequency capacitor applications},\ }\href@noop {} {\bibfield  {journal}
  {\bibinfo  {journal} {Annual Review of Materials Science}\ }\textbf {\bibinfo
  {volume} {28}},\ \bibinfo {pages} {397} (\bibinfo {year} {1998})}\BibitemShut
  {NoStop}%
\bibitem [{\citenamefont {Schlom}\ \emph {et~al.}(2007)\citenamefont {Schlom},
  \citenamefont {Chen}, \citenamefont {Eom}, \citenamefont {Rabe},
  \citenamefont {Streiffer},\ and\ \citenamefont {Triscone}}]{Schlom2007}%
  \BibitemOpen
  \bibfield  {author} {\bibinfo {author} {\bibfnamefont {D.~G.}\ \bibnamefont
  {Schlom}}, \bibinfo {author} {\bibfnamefont {L.~Q.}\ \bibnamefont {Chen}},
  \bibinfo {author} {\bibfnamefont {C.~B.}\ \bibnamefont {Eom}}, \bibinfo
  {author} {\bibfnamefont {K.~M.}\ \bibnamefont {Rabe}}, \bibinfo {author}
  {\bibfnamefont {S.~K.}\ \bibnamefont {Streiffer}},\ and\ \bibinfo {author}
  {\bibfnamefont {J.~M.}\ \bibnamefont {Triscone}},\ }\bibfield  {title}
  {\bibinfo {title} {{Strain tuning of ferroelectric thin films}},\ }\href
  {https://doi.org/10.1146/annurev.matsci.37.061206.113016} {\bibfield
  {journal} {\bibinfo  {journal} {Annual Review of Materials Research}\
  }\textbf {\bibinfo {volume} {37}},\ \bibinfo {pages} {589} (\bibinfo {year}
  {2007})}\BibitemShut {NoStop}%
\bibitem [{\citenamefont {Martin}\ and\ \citenamefont
  {Rappe}(2017)}]{Martin2017}%
  \BibitemOpen
  \bibfield  {author} {\bibinfo {author} {\bibfnamefont {L.~W.}\ \bibnamefont
  {Martin}}\ and\ \bibinfo {author} {\bibfnamefont {A.~M.}\ \bibnamefont
  {Rappe}},\ }\bibfield  {title} {\bibinfo {title} {{Thin-film ferroelectric
  materials and their applications}},\ }\href
  {https://doi.org/10.1038/natrevmats.2016.87} {\bibfield  {journal} {\bibinfo
  {journal} {Nature Reviews Materials}\ }\textbf {\bibinfo {volume} {2}},\
  \bibinfo {pages} {16087} (\bibinfo {year} {2017})}\BibitemShut {NoStop}%
\bibitem [{\citenamefont {Fernandez}\ \emph {et~al.}(2022)\citenamefont
  {Fernandez}, \citenamefont {Acharya}, \citenamefont {Lee}, \citenamefont
  {Schimpf}, \citenamefont {Jiang}, \citenamefont {Lou}, \citenamefont {Tian},\
  and\ \citenamefont {Martin}}]{Fernandez2022}%
  \BibitemOpen
  \bibfield  {author} {\bibinfo {author} {\bibfnamefont {A.}~\bibnamefont
  {Fernandez}}, \bibinfo {author} {\bibfnamefont {M.}~\bibnamefont {Acharya}},
  \bibinfo {author} {\bibfnamefont {H.}~\bibnamefont {Lee}}, \bibinfo {author}
  {\bibfnamefont {J.}~\bibnamefont {Schimpf}}, \bibinfo {author} {\bibfnamefont
  {Y.}~\bibnamefont {Jiang}}, \bibinfo {author} {\bibfnamefont
  {D.}~\bibnamefont {Lou}}, \bibinfo {author} {\bibfnamefont {Z.}~\bibnamefont
  {Tian}},\ and\ \bibinfo {author} {\bibfnamefont {L.~W.}\ \bibnamefont
  {Martin}},\ }\bibfield  {title} {\bibinfo {title} {{Thin‐Film
  Ferroelectrics}},\ }\href {https://doi.org/10.1002/adma.202108841} {\bibfield
   {journal} {\bibinfo  {journal} {Advanced Materials}\ }\textbf {\bibinfo
  {volume} {34}},\ \bibinfo {pages} {1} (\bibinfo {year} {2022})}\BibitemShut
  {NoStop}%
\bibitem [{\citenamefont {Lufaso}\ and\ \citenamefont
  {Woodward}(2001)}]{Lufaso2001}%
  \BibitemOpen
  \bibfield  {author} {\bibinfo {author} {\bibfnamefont {M.~W.}\ \bibnamefont
  {Lufaso}}\ and\ \bibinfo {author} {\bibfnamefont {P.~M.}\ \bibnamefont
  {Woodward}},\ }\bibfield  {title} {\bibinfo {title} {{Prediction of the
  crystal structures of perovskites using the software program SPuDS}},\ }\href
  {https://doi.org/10.1107/S0108768101015282} {\bibfield  {journal} {\bibinfo
  {journal} {Acta Crystallographica Section B: Structural Science}\ }\textbf
  {\bibinfo {volume} {57}},\ \bibinfo {pages} {725} (\bibinfo {year}
  {2001})}\BibitemShut {NoStop}%
\bibitem [{\citenamefont {Woodward}(1997{\natexlab{a}})}]{Woodward1997b}%
  \BibitemOpen
  \bibfield  {author} {\bibinfo {author} {\bibfnamefont {P.~M.}\ \bibnamefont
  {Woodward}},\ }\bibfield  {title} {\bibinfo {title} {Octahedral tilting in
  perovskites. {II.} structure stabilizing forces},\ }\href@noop {} {\bibfield
  {journal} {\bibinfo  {journal} {Acta Crystallographica Section B: Structural
  Science}\ }\textbf {\bibinfo {volume} {53}},\ \bibinfo {pages} {44} (\bibinfo
  {year} {1997}{\natexlab{a}})}\BibitemShut {NoStop}%
\bibitem [{\citenamefont {Woodward}(1997{\natexlab{b}})}]{Woodward1997a}%
  \BibitemOpen
  \bibfield  {author} {\bibinfo {author} {\bibfnamefont {P.~M.}\ \bibnamefont
  {Woodward}},\ }\bibfield  {title} {\bibinfo {title} {Octahedral tilting in
  perovskites. i. geometrical considerations},\ }\href
  {https://doi.org/10.1107/S0108768196010713} {\bibfield  {journal} {\bibinfo
  {journal} {Acta Crystallographica Section B: Structural Science}\ }\textbf
  {\bibinfo {volume} {53}},\ \bibinfo {pages} {32} (\bibinfo {year}
  {1997}{\natexlab{b}})}\BibitemShut {NoStop}%
\bibitem [{\citenamefont {Glazer}(1972)}]{Glazer1972}%
  \BibitemOpen
  \bibfield  {author} {\bibinfo {author} {\bibfnamefont {A.~M.}\ \bibnamefont
  {Glazer}},\ }\bibfield  {title} {\bibinfo {title} {The classification of
  tilted octahedra in perovskites},\ }\href
  {https://doi.org/10.1107/S0567740872007976} {\bibfield  {journal} {\bibinfo
  {journal} {Acta Crystallographica Section B Structural Crystallography and
  Crystal Chemistry}\ }\textbf {\bibinfo {volume} {28}},\ \bibinfo {pages}
  {3384} (\bibinfo {year} {1972})}\BibitemShut {NoStop}%
\bibitem [{\citenamefont {Howard}\ and\ \citenamefont
  {Stokes}(1998)}]{Howard1998}%
  \BibitemOpen
  \bibfield  {author} {\bibinfo {author} {\bibfnamefont {C.~J.}\ \bibnamefont
  {Howard}}\ and\ \bibinfo {author} {\bibfnamefont {H.~T.}\ \bibnamefont
  {Stokes}},\ }\bibfield  {title} {\bibinfo {title} {Group-theoretical analysis
  of octahedral tilting in perovskites},\ }\href
  {https://doi.org/10.1107/S0108768198004200} {\bibfield  {journal} {\bibinfo
  {journal} {Acta Crystallographica Section B Structural Science}\ }\textbf
  {\bibinfo {volume} {54}},\ \bibinfo {pages} {782} (\bibinfo {year}
  {1998})}\BibitemShut {NoStop}%
\bibitem [{\citenamefont {Wang}\ \emph {et~al.}(2022)\citenamefont {Wang},
  \citenamefont {Gautreau}, \citenamefont {Birol},\ and\ \citenamefont
  {Fernandes}}]{Wang2022RTO}%
  \BibitemOpen
  \bibfield  {author} {\bibinfo {author} {\bibfnamefont {Z.}~\bibnamefont
  {Wang}}, \bibinfo {author} {\bibfnamefont {D.}~\bibnamefont {Gautreau}},
  \bibinfo {author} {\bibfnamefont {T.}~\bibnamefont {Birol}},\ and\ \bibinfo
  {author} {\bibfnamefont {R.~M.}\ \bibnamefont {Fernandes}},\ }\bibfield
  {title} {\bibinfo {title} {Strain-tunable metamagnetic critical endpoint in
  mott insulating rare-earth titanates},\ }\bibfield  {journal} {\bibinfo
  {journal} {Physical Review B}\ }\textbf {\bibinfo {volume} {105}},\ \href
  {https://doi.org/10.1103/PhysRevB.105.144404} {10.1103/PhysRevB.105.144404}
  (\bibinfo {year} {2022})\BibitemShut {NoStop}%
\bibitem [{\citenamefont {Najev}\ \emph {et~al.}(2022)\citenamefont {Najev},
  \citenamefont {Hameed}, \citenamefont {Gautreau}, \citenamefont {Wang},
  \citenamefont {Joe}, \citenamefont {PoŽek}, \citenamefont {Birol},
  \citenamefont {Fernandes}, \citenamefont {Greven},\ and\ \citenamefont
  {Pelc}}]{Najev2022}%
  \BibitemOpen
  \bibfield  {author} {\bibinfo {author} {\bibfnamefont {A.}~\bibnamefont
  {Najev}}, \bibinfo {author} {\bibfnamefont {S.}~\bibnamefont {Hameed}},
  \bibinfo {author} {\bibfnamefont {D.}~\bibnamefont {Gautreau}}, \bibinfo
  {author} {\bibfnamefont {Z.}~\bibnamefont {Wang}}, \bibinfo {author}
  {\bibfnamefont {J.}~\bibnamefont {Joe}}, \bibinfo {author} {\bibfnamefont
  {M.}~\bibnamefont {PoŽek}}, \bibinfo {author} {\bibfnamefont
  {T.}~\bibnamefont {Birol}}, \bibinfo {author} {\bibfnamefont {R.~M.}\
  \bibnamefont {Fernandes}}, \bibinfo {author} {\bibfnamefont {M.}~\bibnamefont
  {Greven}},\ and\ \bibinfo {author} {\bibfnamefont {D.}~\bibnamefont {Pelc}},\
  }\bibfield  {title} {\bibinfo {title} {Uniaxial strain control of bulk
  ferromagnetism in rare-earth titanates},\ }\bibfield  {journal} {\bibinfo
  {journal} {Physical Review Letters}\ }\textbf {\bibinfo {volume} {128}},\
  \href {https://doi.org/10.1103/PhysRevLett.128.167201}
  {10.1103/PhysRevLett.128.167201} (\bibinfo {year} {2022})\BibitemShut
  {NoStop}%
\bibitem [{\citenamefont {Choquette}\ \emph {et~al.}(2016)\citenamefont
  {Choquette}, \citenamefont {Smith}, \citenamefont {Sichel-Tissot},
  \citenamefont {Moon}, \citenamefont {Scafetta}, \citenamefont {Gennaro},
  \citenamefont {Granozio}, \citenamefont {Karapetrova},\ and\ \citenamefont
  {May}}]{Choquette2016}%
  \BibitemOpen
  \bibfield  {author} {\bibinfo {author} {\bibfnamefont {A.~K.}\ \bibnamefont
  {Choquette}}, \bibinfo {author} {\bibfnamefont {C.~R.}\ \bibnamefont
  {Smith}}, \bibinfo {author} {\bibfnamefont {R.~J.}\ \bibnamefont
  {Sichel-Tissot}}, \bibinfo {author} {\bibfnamefont {E.~J.}\ \bibnamefont
  {Moon}}, \bibinfo {author} {\bibfnamefont {M.~D.}\ \bibnamefont {Scafetta}},
  \bibinfo {author} {\bibfnamefont {E.~D.}\ \bibnamefont {Gennaro}}, \bibinfo
  {author} {\bibfnamefont {F.~M.}\ \bibnamefont {Granozio}}, \bibinfo {author}
  {\bibfnamefont {E.}~\bibnamefont {Karapetrova}},\ and\ \bibinfo {author}
  {\bibfnamefont {S.~J.}\ \bibnamefont {May}},\ }\bibfield  {title} {\bibinfo
  {title} {{Octahedral rotation patterns in strained {EuFeO$_3$} and other Pbnm
  perovskite films: Implications for hybrid improper ferroelectricity}},\
  }\bibfield  {journal} {\bibinfo  {journal} {Physical Review B}\ }\textbf
  {\bibinfo {volume} {94}},\ \href {https://doi.org/10.1103/PhysRevB.94.024105}
  {10.1103/PhysRevB.94.024105} (\bibinfo {year} {2016})\BibitemShut {NoStop}%
\bibitem [{\citenamefont {Kay}\ and\ \citenamefont {Bailey}(1957)}]{Kay1957}%
  \BibitemOpen
  \bibfield  {author} {\bibinfo {author} {\bibfnamefont {H.~F.}\ \bibnamefont
  {Kay}}\ and\ \bibinfo {author} {\bibfnamefont {P.~C.}\ \bibnamefont
  {Bailey}},\ }\bibfield  {title} {\bibinfo {title} {Structure and properties
  of {CaTiO$_3$}},\ }\href {https://doi.org/10.1107/S0365110X57000675}
  {\bibfield  {journal} {\bibinfo  {journal} {Acta Crystallographica}\ }\textbf
  {\bibinfo {volume} {10}},\ \bibinfo {pages} {219} (\bibinfo {year}
  {1957})}\BibitemShut {NoStop}%
\bibitem [{\citenamefont {Biegalski}\ \emph {et~al.}(2015)\citenamefont
  {Biegalski}, \citenamefont {Qiao}, \citenamefont {Gu}, \citenamefont {Mehta},
  \citenamefont {He}, \citenamefont {Takamura}, \citenamefont {Borisevich},\
  and\ \citenamefont {Chen}}]{Biegalski2015}%
  \BibitemOpen
  \bibfield  {author} {\bibinfo {author} {\bibfnamefont {M.~D.}\ \bibnamefont
  {Biegalski}}, \bibinfo {author} {\bibfnamefont {L.}~\bibnamefont {Qiao}},
  \bibinfo {author} {\bibfnamefont {Y.}~\bibnamefont {Gu}}, \bibinfo {author}
  {\bibfnamefont {A.}~\bibnamefont {Mehta}}, \bibinfo {author} {\bibfnamefont
  {Q.}~\bibnamefont {He}}, \bibinfo {author} {\bibfnamefont {Y.}~\bibnamefont
  {Takamura}}, \bibinfo {author} {\bibfnamefont {A.}~\bibnamefont
  {Borisevich}},\ and\ \bibinfo {author} {\bibfnamefont {L.~Q.}\ \bibnamefont
  {Chen}},\ }\bibfield  {title} {\bibinfo {title} {Impact of symmetry on the
  ferroelectric properties of {CaTiO$_3$} thin films},\ }\bibfield  {journal}
  {\bibinfo  {journal} {Applied Physics Letters}\ }\textbf {\bibinfo {volume}
  {106}},\ \href {https://doi.org/10.1063/1.4918805} {10.1063/1.4918805}
  (\bibinfo {year} {2015})\BibitemShut {NoStop}%
\bibitem [{\citenamefont {Wang}\ \emph {et~al.}(2018)\citenamefont {Wang},
  \citenamefont {Prakash}, \citenamefont {Dong}, \citenamefont {Truttmann},
  \citenamefont {Bucsek}, \citenamefont {James}, \citenamefont {Fong},
  \citenamefont {Kim}, \citenamefont {Ryan}, \citenamefont {Zhou},
  \citenamefont {Birol},\ and\ \citenamefont {Jalan}}]{Wang2018}%
  \BibitemOpen
  \bibfield  {author} {\bibinfo {author} {\bibfnamefont {T.}~\bibnamefont
  {Wang}}, \bibinfo {author} {\bibfnamefont {A.}~\bibnamefont {Prakash}},
  \bibinfo {author} {\bibfnamefont {Y.}~\bibnamefont {Dong}}, \bibinfo {author}
  {\bibfnamefont {T.}~\bibnamefont {Truttmann}}, \bibinfo {author}
  {\bibfnamefont {A.}~\bibnamefont {Bucsek}}, \bibinfo {author} {\bibfnamefont
  {R.}~\bibnamefont {James}}, \bibinfo {author} {\bibfnamefont {D.~D.}\
  \bibnamefont {Fong}}, \bibinfo {author} {\bibfnamefont {J.-W.}\ \bibnamefont
  {Kim}}, \bibinfo {author} {\bibfnamefont {P.~J.}\ \bibnamefont {Ryan}},
  \bibinfo {author} {\bibfnamefont {H.}~\bibnamefont {Zhou}}, \bibinfo {author}
  {\bibfnamefont {T.}~\bibnamefont {Birol}},\ and\ \bibinfo {author}
  {\bibfnamefont {B.}~\bibnamefont {Jalan}},\ }\bibfield  {title} {\bibinfo
  {title} {Engineering {SrSnO$_3$} phases and electron mobility at room
  temperature using epitaxial strain},\ }\href
  {https://doi.org/10.1021/acsami.8b16592} {\bibfield  {journal} {\bibinfo
  {journal} {{ACS Applied Materials \& Interfaces}}\ }\textbf {\bibinfo
  {volume} {10}},\ \bibinfo {pages} {43802} (\bibinfo {year}
  {2018})}\BibitemShut {NoStop}%
\bibitem [{\citenamefont {Okamoto}(2013)}]{Okamoto2013}%
  \BibitemOpen
  \bibfield  {author} {\bibinfo {author} {\bibfnamefont {S.}~\bibnamefont
  {Okamoto}},\ }\bibfield  {title} {\bibinfo {title} {Doped mott insulators in
  (111) bilayers of perovskite transition-metal oxides with a strong spin-orbit
  coupling},\ }\bibfield  {journal} {\bibinfo  {journal} {Physical Review
  Letters}\ }\textbf {\bibinfo {volume} {110}},\ \href
  {https://doi.org/10.1103/PhysRevLett.110.066403}
  {10.1103/PhysRevLett.110.066403} (\bibinfo {year} {2013})\BibitemShut
  {NoStop}%
\bibitem [{\citenamefont {Rüegg}\ and\ \citenamefont
  {Fiete}(2011)}]{Ruegg2011}%
  \BibitemOpen
  \bibfield  {author} {\bibinfo {author} {\bibfnamefont {A.}~\bibnamefont
  {Rüegg}}\ and\ \bibinfo {author} {\bibfnamefont {G.~A.}\ \bibnamefont
  {Fiete}},\ }\bibfield  {title} {\bibinfo {title} {Topological insulators from
  complex orbital order in transition-metal oxides heterostructures},\ }\href
  {https://doi.org/10.1103/PhysRevB.84.201103} {\bibfield  {journal} {\bibinfo
  {journal} {Physical Review B}\ }\textbf {\bibinfo {volume} {84}},\ \bibinfo
  {pages} {201103} (\bibinfo {year} {2011})}\BibitemShut {NoStop}%
\bibitem [{\citenamefont {Chakraverty}\ \emph {et~al.}(2010)\citenamefont
  {Chakraverty}, \citenamefont {Ohtomo}, \citenamefont {Okude}, \citenamefont
  {Ueno},\ and\ \citenamefont {Kawasaki}}]{Chakraverty2010}%
  \BibitemOpen
  \bibfield  {author} {\bibinfo {author} {\bibfnamefont {S.}~\bibnamefont
  {Chakraverty}}, \bibinfo {author} {\bibfnamefont {A.}~\bibnamefont {Ohtomo}},
  \bibinfo {author} {\bibfnamefont {M.}~\bibnamefont {Okude}}, \bibinfo
  {author} {\bibfnamefont {K.}~\bibnamefont {Ueno}},\ and\ \bibinfo {author}
  {\bibfnamefont {M.}~\bibnamefont {Kawasaki}},\ }\bibfield  {title} {\bibinfo
  {title} {{Epitaxial structure of (001)- and (111)-oriented perovskite ferrate
  films grown by pulsed-laser deposition}},\ }\href
  {https://doi.org/10.1021/cg901355c} {\bibfield  {journal} {\bibinfo
  {journal} {Crystal Growth and Design}\ }\textbf {\bibinfo {volume} {10}},\
  \bibinfo {pages} {1725} (\bibinfo {year} {2010})}\BibitemShut {NoStop}%
\bibitem [{\citenamefont {Tang}\ \emph {et~al.}(2019)\citenamefont {Tang},
  \citenamefont {Zhu}, \citenamefont {Ma}, \citenamefont {Hong}, \citenamefont
  {Wang}, \citenamefont {Wang}, \citenamefont {Xu}, \citenamefont {Liu},
  \citenamefont {Wu}, \citenamefont {Chen}, \citenamefont {Huang},
  \citenamefont {Chen}, \citenamefont {Chen}, \citenamefont {Wu},\ and\
  \citenamefont {Pennycook}}]{Tang2019PTO}%
  \BibitemOpen
  \bibfield  {author} {\bibinfo {author} {\bibfnamefont {Y.}~\bibnamefont
  {Tang}}, \bibinfo {author} {\bibfnamefont {Y.}~\bibnamefont {Zhu}}, \bibinfo
  {author} {\bibfnamefont {X.}~\bibnamefont {Ma}}, \bibinfo {author}
  {\bibfnamefont {Z.}~\bibnamefont {Hong}}, \bibinfo {author} {\bibfnamefont
  {Y.}~\bibnamefont {Wang}}, \bibinfo {author} {\bibfnamefont {W.}~\bibnamefont
  {Wang}}, \bibinfo {author} {\bibfnamefont {Y.}~\bibnamefont {Xu}}, \bibinfo
  {author} {\bibfnamefont {Y.}~\bibnamefont {Liu}}, \bibinfo {author}
  {\bibfnamefont {B.}~\bibnamefont {Wu}}, \bibinfo {author} {\bibfnamefont
  {L.}~\bibnamefont {Chen}}, \bibinfo {author} {\bibfnamefont {C.}~\bibnamefont
  {Huang}}, \bibinfo {author} {\bibfnamefont {L.}~\bibnamefont {Chen}},
  \bibinfo {author} {\bibfnamefont {Z.}~\bibnamefont {Chen}}, \bibinfo {author}
  {\bibfnamefont {H.}~\bibnamefont {Wu}},\ and\ \bibinfo {author}
  {\bibfnamefont {S.~J.}\ \bibnamefont {Pennycook}},\ }\bibfield  {title}
  {\bibinfo {title} {{A Coherently Strained Monoclinic [111] {PbTiO$_3$} Film
  Exhibiting Zero Poisson's Ratio State}},\ }\href
  {https://doi.org/10.1002/adfm.201901687} {\bibfield  {journal} {\bibinfo
  {journal} {Advanced Functional Materials}\ }\textbf {\bibinfo {volume}
  {29}},\ \bibinfo {pages} {1} (\bibinfo {year} {2019})}\BibitemShut {NoStop}%
\bibitem [{\citenamefont {Liang}\ \emph {et~al.}(2015)\citenamefont {Liang},
  \citenamefont {Li}, \citenamefont {Zhang}, \citenamefont {Lin}, \citenamefont
  {Li}, \citenamefont {Yao}, \citenamefont {Li}, \citenamefont {Yang},\ and\
  \citenamefont {Guo}}]{Liang2015}%
  \BibitemOpen
  \bibfield  {author} {\bibinfo {author} {\bibfnamefont {Y.}~\bibnamefont
  {Liang}}, \bibinfo {author} {\bibfnamefont {W.}~\bibnamefont {Li}}, \bibinfo
  {author} {\bibfnamefont {S.}~\bibnamefont {Zhang}}, \bibinfo {author}
  {\bibfnamefont {C.}~\bibnamefont {Lin}}, \bibinfo {author} {\bibfnamefont
  {C.}~\bibnamefont {Li}}, \bibinfo {author} {\bibfnamefont {Y.}~\bibnamefont
  {Yao}}, \bibinfo {author} {\bibfnamefont {Y.}~\bibnamefont {Li}}, \bibinfo
  {author} {\bibfnamefont {H.}~\bibnamefont {Yang}},\ and\ \bibinfo {author}
  {\bibfnamefont {J.}~\bibnamefont {Guo}},\ }\bibfield  {title} {\bibinfo
  {title} {{Homoepitaxial {SrTiO$_3$} (111) Film with High Dielectric
  Performance and Atomically Well-Defined Surface}},\ }\href
  {https://doi.org/10.1038/srep10634} {\bibfield  {journal} {\bibinfo
  {journal} {Scientific Reports}\ }\textbf {\bibinfo {volume} {5}},\ \bibinfo
  {pages} {1} (\bibinfo {year} {2015})}\BibitemShut {NoStop}%
\bibitem [{\citenamefont {Roth}\ \emph {et~al.}(2021)\citenamefont {Roth},
  \citenamefont {Kuznetsova}, \citenamefont {Miao}, \citenamefont
  {Pogrebnyakov}, \citenamefont {Alem},\ and\ \citenamefont
  {Engel-Herbert}}]{Roth2021SVO111}%
  \BibitemOpen
  \bibfield  {author} {\bibinfo {author} {\bibfnamefont {J.}~\bibnamefont
  {Roth}}, \bibinfo {author} {\bibfnamefont {T.}~\bibnamefont {Kuznetsova}},
  \bibinfo {author} {\bibfnamefont {L.}~\bibnamefont {Miao}}, \bibinfo {author}
  {\bibfnamefont {A.}~\bibnamefont {Pogrebnyakov}}, \bibinfo {author}
  {\bibfnamefont {N.}~\bibnamefont {Alem}},\ and\ \bibinfo {author}
  {\bibfnamefont {R.}~\bibnamefont {Engel-Herbert}},\ }\bibfield  {title}
  {\bibinfo {title} {{Self-regulated growth of [111]-oriented perovskite oxide
  films using hybrid molecular beam epitaxy}},\ }\bibfield  {journal} {\bibinfo
   {journal} {APL Materials}\ }\textbf {\bibinfo {volume} {9}},\ \href
  {https://doi.org/10.1063/5.0040047} {10.1063/5.0040047} (\bibinfo {year}
  {2021})\BibitemShut {NoStop}%
\bibitem [{\citenamefont {Middey}\ \emph {et~al.}(2014)\citenamefont {Middey},
  \citenamefont {Rivero}, \citenamefont {Meyers}, \citenamefont {Kareev},
  \citenamefont {Liu}, \citenamefont {Cao}, \citenamefont {Freeland},
  \citenamefont {Barraza-Lopez},\ and\ \citenamefont
  {Chakhalian}}]{Middey2014}%
  \BibitemOpen
  \bibfield  {author} {\bibinfo {author} {\bibfnamefont {S.}~\bibnamefont
  {Middey}}, \bibinfo {author} {\bibfnamefont {P.}~\bibnamefont {Rivero}},
  \bibinfo {author} {\bibfnamefont {D.}~\bibnamefont {Meyers}}, \bibinfo
  {author} {\bibfnamefont {M.}~\bibnamefont {Kareev}}, \bibinfo {author}
  {\bibfnamefont {X.}~\bibnamefont {Liu}}, \bibinfo {author} {\bibfnamefont
  {Y.}~\bibnamefont {Cao}}, \bibinfo {author} {\bibfnamefont {J.~W.}\
  \bibnamefont {Freeland}}, \bibinfo {author} {\bibfnamefont {S.}~\bibnamefont
  {Barraza-Lopez}},\ and\ \bibinfo {author} {\bibfnamefont {J.}~\bibnamefont
  {Chakhalian}},\ }\bibfield  {title} {\bibinfo {title} {Polarity compensation
  in ultra-thin films of complex oxides: The case of a perovskite nickelate},\
  }\bibfield  {journal} {\bibinfo  {journal} {Scientific Reports}\ }\textbf
  {\bibinfo {volume} {4}},\ \href {https://doi.org/10.1038/srep06819}
  {10.1038/srep06819} (\bibinfo {year} {2014})\BibitemShut {NoStop}%
\bibitem [{\citenamefont {Middey}\ \emph {et~al.}(2016)\citenamefont {Middey},
  \citenamefont {Meyers}, \citenamefont {Doennig}, \citenamefont {Kareev},
  \citenamefont {Liu}, \citenamefont {Cao}, \citenamefont {Yang}, \citenamefont
  {Shi}, \citenamefont {Gu}, \citenamefont {Ryan}, \citenamefont {Pentcheva},
  \citenamefont {Freeland},\ and\ \citenamefont {Chakhalian}}]{Middey2016}%
  \BibitemOpen
  \bibfield  {author} {\bibinfo {author} {\bibfnamefont {S.}~\bibnamefont
  {Middey}}, \bibinfo {author} {\bibfnamefont {D.}~\bibnamefont {Meyers}},
  \bibinfo {author} {\bibfnamefont {D.}~\bibnamefont {Doennig}}, \bibinfo
  {author} {\bibfnamefont {M.}~\bibnamefont {Kareev}}, \bibinfo {author}
  {\bibfnamefont {X.}~\bibnamefont {Liu}}, \bibinfo {author} {\bibfnamefont
  {Y.}~\bibnamefont {Cao}}, \bibinfo {author} {\bibfnamefont {Z.}~\bibnamefont
  {Yang}}, \bibinfo {author} {\bibfnamefont {J.}~\bibnamefont {Shi}}, \bibinfo
  {author} {\bibfnamefont {L.}~\bibnamefont {Gu}}, \bibinfo {author}
  {\bibfnamefont {P.~J.}\ \bibnamefont {Ryan}}, \bibinfo {author}
  {\bibfnamefont {R.}~\bibnamefont {Pentcheva}}, \bibinfo {author}
  {\bibfnamefont {J.~W.}\ \bibnamefont {Freeland}},\ and\ \bibinfo {author}
  {\bibfnamefont {J.}~\bibnamefont {Chakhalian}},\ }\bibfield  {title}
  {\bibinfo {title} {Mott electrons in an artificial graphenelike crystal of
  rare-earth nickelate},\ }\bibfield  {journal} {\bibinfo  {journal} {Physical
  Review Letters}\ }\textbf {\bibinfo {volume} {116}},\ \href
  {https://doi.org/10.1103/PhysRevLett.116.056801}
  {10.1103/PhysRevLett.116.056801} (\bibinfo {year} {2016})\BibitemShut
  {NoStop}%
\bibitem [{\citenamefont {Gibert}\ \emph {et~al.}(2016)\citenamefont {Gibert},
  \citenamefont {Viret}, \citenamefont {Zubko}, \citenamefont {Jaouen},
  \citenamefont {Tonnerre}, \citenamefont {Torres-Pardo}, \citenamefont
  {Catalano}, \citenamefont {Gloter}, \citenamefont {Stéphan},\ and\
  \citenamefont {Triscone}}]{Gibert2016}%
  \BibitemOpen
  \bibfield  {author} {\bibinfo {author} {\bibfnamefont {M.}~\bibnamefont
  {Gibert}}, \bibinfo {author} {\bibfnamefont {M.}~\bibnamefont {Viret}},
  \bibinfo {author} {\bibfnamefont {P.}~\bibnamefont {Zubko}}, \bibinfo
  {author} {\bibfnamefont {N.}~\bibnamefont {Jaouen}}, \bibinfo {author}
  {\bibfnamefont {J.~M.}\ \bibnamefont {Tonnerre}}, \bibinfo {author}
  {\bibfnamefont {A.}~\bibnamefont {Torres-Pardo}}, \bibinfo {author}
  {\bibfnamefont {S.}~\bibnamefont {Catalano}}, \bibinfo {author}
  {\bibfnamefont {A.}~\bibnamefont {Gloter}}, \bibinfo {author} {\bibfnamefont
  {O.}~\bibnamefont {Stéphan}},\ and\ \bibinfo {author} {\bibfnamefont
  {J.~M.}\ \bibnamefont {Triscone}},\ }\bibfield  {title} {\bibinfo {title}
  {Interlayer coupling through a dimensionality-induced magnetic state},\
  }\bibfield  {journal} {\bibinfo  {journal} {Nature Communications}\ }\textbf
  {\bibinfo {volume} {7}},\ \href {https://doi.org/10.1038/ncomms11227}
  {10.1038/ncomms11227} (\bibinfo {year} {2016})\BibitemShut {NoStop}%
\bibitem [{\citenamefont {Kim}\ \emph {et~al.}(2023)\citenamefont {Kim},
  \citenamefont {Yu}, \citenamefont {Lee}, \citenamefont {Shang}, \citenamefont
  {Kim}, \citenamefont {Pal}, \citenamefont {Seo}, \citenamefont {Campbell},
  \citenamefont {Eom}, \citenamefont {Ramachandran}, \citenamefont {Rzchowski},
  \citenamefont {Oh}, \citenamefont {Choi}, \citenamefont {Liu}, \citenamefont
  {Levy},\ and\ \citenamefont {Eom}}]{Kim2023KTO}%
  \BibitemOpen
  \bibfield  {author} {\bibinfo {author} {\bibfnamefont {J.}~\bibnamefont
  {Kim}}, \bibinfo {author} {\bibfnamefont {M.}~\bibnamefont {Yu}}, \bibinfo
  {author} {\bibfnamefont {J.-W.}\ \bibnamefont {Lee}}, \bibinfo {author}
  {\bibfnamefont {S.-L.}\ \bibnamefont {Shang}}, \bibinfo {author}
  {\bibfnamefont {G.-Y.}\ \bibnamefont {Kim}}, \bibinfo {author} {\bibfnamefont
  {P.}~\bibnamefont {Pal}}, \bibinfo {author} {\bibfnamefont {J.}~\bibnamefont
  {Seo}}, \bibinfo {author} {\bibfnamefont {N.}~\bibnamefont {Campbell}},
  \bibinfo {author} {\bibfnamefont {K.}~\bibnamefont {Eom}}, \bibinfo {author}
  {\bibfnamefont {R.}~\bibnamefont {Ramachandran}}, \bibinfo {author}
  {\bibfnamefont {M.~S.}\ \bibnamefont {Rzchowski}}, \bibinfo {author}
  {\bibfnamefont {S.~H.}\ \bibnamefont {Oh}}, \bibinfo {author} {\bibfnamefont
  {S.-Y.}\ \bibnamefont {Choi}}, \bibinfo {author} {\bibfnamefont {Z.-K.}\
  \bibnamefont {Liu}}, \bibinfo {author} {\bibfnamefont {J.}~\bibnamefont
  {Levy}},\ and\ \bibinfo {author} {\bibfnamefont {C.-B.}\ \bibnamefont
  {Eom}},\ }\href@noop {} {\bibinfo {title} {Electronic-grade epitaxial (111)
  {KTaO$_3$} heterostructures}} (\bibinfo {year} {2023}),\ \Eprint
  {https://arxiv.org/abs/2308.13180} {arXiv:2308.13180 [cond-mat.mtrl-sci]}
  \BibitemShut {NoStop}%
\bibitem [{\citenamefont {van Roosmalen}\ \emph {et~al.}(1992)\citenamefont
  {van Roosmalen}, \citenamefont {van Vlaanderen},\ and\ \citenamefont
  {Cordfunke}}]{Roosmalen1992}%
  \BibitemOpen
  \bibfield  {author} {\bibinfo {author} {\bibfnamefont {J.}~\bibnamefont {van
  Roosmalen}}, \bibinfo {author} {\bibfnamefont {P.}~\bibnamefont {van
  Vlaanderen}},\ and\ \bibinfo {author} {\bibfnamefont {E.}~\bibnamefont
  {Cordfunke}},\ }\bibfield  {title} {\bibinfo {title} {On the structure of
  {SrZrO$_3$}},\ }\href {https://doi.org/10.1016/0022-4596(92)90200-F}
  {\bibfield  {journal} {\bibinfo  {journal} {Journal of Solid State
  Chemistry}\ }\textbf {\bibinfo {volume} {101}},\ \bibinfo {pages} {59}
  (\bibinfo {year} {1992})}\BibitemShut {NoStop}%
\bibitem [{\citenamefont {Glerup}\ \emph {et~al.}(2005)\citenamefont {Glerup},
  \citenamefont {Knight},\ and\ \citenamefont {Poulsen}}]{Glerup2005}%
  \BibitemOpen
  \bibfield  {author} {\bibinfo {author} {\bibfnamefont {M.}~\bibnamefont
  {Glerup}}, \bibinfo {author} {\bibfnamefont {K.~S.}\ \bibnamefont {Knight}},\
  and\ \bibinfo {author} {\bibfnamefont {F.~W.}\ \bibnamefont {Poulsen}},\
  }\bibfield  {title} {\bibinfo {title} {High temperature structural phase
  transitions in {SrSnO$_3$} perovskite},\ }\href
  {https://doi.org/10.1016/j.materresbull.2004.11.004} {\bibfield  {journal}
  {\bibinfo  {journal} {Materials Research Bulletin}\ }\textbf {\bibinfo
  {volume} {40}},\ \bibinfo {pages} {507} (\bibinfo {year} {2005})}\BibitemShut
  {NoStop}%
\bibitem [{\citenamefont {Yashima}\ and\ \citenamefont
  {Ali}(2009)}]{Yashima2009}%
  \BibitemOpen
  \bibfield  {author} {\bibinfo {author} {\bibfnamefont {M.}~\bibnamefont
  {Yashima}}\ and\ \bibinfo {author} {\bibfnamefont {R.}~\bibnamefont {Ali}},\
  }\bibfield  {title} {\bibinfo {title} {Structural phase transition and
  octahedral tilting in the calcium titanate perovskite {CaTiO$_3$}},\ }\href
  {https://doi.org/10.1016/j.ssi.2008.11.019} {\bibfield  {journal} {\bibinfo
  {journal} {Solid State Ionics}\ }\textbf {\bibinfo {volume} {180}},\ \bibinfo
  {pages} {120} (\bibinfo {year} {2009})}\BibitemShut {NoStop}%
\bibitem [{\citenamefont {Kennedy}\ \emph {et~al.}(1999)\citenamefont
  {Kennedy}, \citenamefont {Howard},\ and\ \citenamefont
  {Chakoumakos}}]{Kennedy1999}%
  \BibitemOpen
  \bibfield  {author} {\bibinfo {author} {\bibfnamefont {B.~J.}\ \bibnamefont
  {Kennedy}}, \bibinfo {author} {\bibfnamefont {C.~J.}\ \bibnamefont
  {Howard}},\ and\ \bibinfo {author} {\bibfnamefont {B.~C.}\ \bibnamefont
  {Chakoumakos}},\ }\bibfield  {title} {\bibinfo {title} {High-temperature
  phase transitions in {SrZrO$_3$}},\ }\href
  {https://doi.org/10.1103/PhysRevB.59.4023} {\bibfield  {journal} {\bibinfo
  {journal} {Physical Review B}\ }\textbf {\bibinfo {volume} {59}},\ \bibinfo
  {pages} {4023} (\bibinfo {year} {1999})}\BibitemShut {NoStop}%
\bibitem [{\citenamefont {Kresse}\ and\ \citenamefont
  {Furthmüller}(1996)}]{Kresse199607}%
  \BibitemOpen
  \bibfield  {author} {\bibinfo {author} {\bibfnamefont {G.}~\bibnamefont
  {Kresse}}\ and\ \bibinfo {author} {\bibfnamefont {J.}~\bibnamefont
  {Furthmüller}},\ }\bibfield  {title} {\bibinfo {title} {Efficiency of
  ab-initio total energy calculations for metals and semiconductors using a
  plane-wave basis set},\ }\href@noop {} {\bibfield  {journal} {\bibinfo
  {journal} {Computational Materials Science}\ }\textbf {\bibinfo {volume}
  {6}},\ \bibinfo {pages} {15} (\bibinfo {year} {1996})}\BibitemShut {NoStop}%
\bibitem [{\citenamefont {Kresse}\ and\ \citenamefont
  {Furthm\"uller}(1996)}]{Kresse199610}%
  \BibitemOpen
  \bibfield  {author} {\bibinfo {author} {\bibfnamefont {G.}~\bibnamefont
  {Kresse}}\ and\ \bibinfo {author} {\bibfnamefont {J.}~\bibnamefont
  {Furthm\"uller}},\ }\bibfield  {title} {\bibinfo {title} {Efficient iterative
  schemes for ab initio total-energy calculations using a plane-wave basis
  set},\ }\href {https://doi.org/10.1103/PhysRevB.54.11169} {\bibfield
  {journal} {\bibinfo  {journal} {Phys. Rev. B}\ }\textbf {\bibinfo {volume}
  {54}},\ \bibinfo {pages} {11169} (\bibinfo {year} {1996})}\BibitemShut
  {NoStop}%
\bibitem [{\citenamefont {Bl{\"o}chl}(1994)}]{Blochl1994}%
  \BibitemOpen
  \bibfield  {author} {\bibinfo {author} {\bibfnamefont {P.~E.}\ \bibnamefont
  {Bl{\"o}chl}},\ }\bibfield  {title} {\bibinfo {title} {Projector
  augmented-wave method},\ }\href@noop {} {\bibfield  {journal} {\bibinfo
  {journal} {Physical review B}\ }\textbf {\bibinfo {volume} {50}},\ \bibinfo
  {pages} {17953} (\bibinfo {year} {1994})}\BibitemShut {NoStop}%
\bibitem [{\citenamefont {Perdew}\ \emph {et~al.}(2008)\citenamefont {Perdew},
  \citenamefont {Ruzsinszky}, \citenamefont {Csonka}, \citenamefont {Vydrov},
  \citenamefont {Scuseria}, \citenamefont {Constantin}, \citenamefont {Zhou},\
  and\ \citenamefont {Burke}}]{Perdew2009}%
  \BibitemOpen
  \bibfield  {author} {\bibinfo {author} {\bibfnamefont {J.~P.}\ \bibnamefont
  {Perdew}}, \bibinfo {author} {\bibfnamefont {A.}~\bibnamefont {Ruzsinszky}},
  \bibinfo {author} {\bibfnamefont {G.~I.}\ \bibnamefont {Csonka}}, \bibinfo
  {author} {\bibfnamefont {O.~A.}\ \bibnamefont {Vydrov}}, \bibinfo {author}
  {\bibfnamefont {G.~E.}\ \bibnamefont {Scuseria}}, \bibinfo {author}
  {\bibfnamefont {L.~A.}\ \bibnamefont {Constantin}}, \bibinfo {author}
  {\bibfnamefont {X.}~\bibnamefont {Zhou}},\ and\ \bibinfo {author}
  {\bibfnamefont {K.}~\bibnamefont {Burke}},\ }\bibfield  {title} {\bibinfo
  {title} {Restoring the density-gradient expansion for exchange in solids and
  surfaces},\ }\href {https://doi.org/10.1103/PhysRevLett.100.136406}
  {\bibfield  {journal} {\bibinfo  {journal} {Phys. Rev. Lett.}\ }\textbf
  {\bibinfo {volume} {100}},\ \bibinfo {pages} {136406} (\bibinfo {year}
  {2008})}\BibitemShut {NoStop}%
\bibitem [{\citenamefont {{H. T. Stokes, D. M. Hatch, and B. J.
  Campbell}}({\natexlab{a}})}]{FINDSYM}%
  \BibitemOpen
  \bibfield  {author} {\bibinfo {author} {\bibnamefont {{H. T. Stokes, D. M.
  Hatch, and B. J. Campbell}}},\ }\href@noop {} {\bibinfo {title} {{FINDSYM,
  ISOTROPY Software Suite, iso.byu.edu.}}},\ \bibinfo {howpublished}
  {\url{https://stokes.byu.edu/iso/findsym.php}} ({\natexlab{a}}),\ \bibinfo
  {note} {accessed: 2023-06-01}\BibitemShut {NoStop}%
\bibitem [{\citenamefont {Stokes}\ and\ \citenamefont
  {Hatch}(2005)}]{Stokes2005}%
  \BibitemOpen
  \bibfield  {author} {\bibinfo {author} {\bibfnamefont {H.~T.}\ \bibnamefont
  {Stokes}}\ and\ \bibinfo {author} {\bibfnamefont {D.~M.}\ \bibnamefont
  {Hatch}},\ }\bibfield  {title} {\bibinfo {title} {Findsym: program for
  identifying the space-group symmetry of a crystal},\ }\href@noop {}
  {\bibfield  {journal} {\bibinfo  {journal} {Journal of Applied
  Crystallography}\ }\textbf {\bibinfo {volume} {38}},\ \bibinfo {pages} {237}
  (\bibinfo {year} {2005})}\BibitemShut {NoStop}%
\bibitem [{\citenamefont {{H. T. Stokes, D. M. Hatch, and B. J.
  Campbell}}({\natexlab{b}})}]{ISODISTORT}%
  \BibitemOpen
  \bibfield  {author} {\bibinfo {author} {\bibnamefont {{H. T. Stokes, D. M.
  Hatch, and B. J. Campbell}}},\ }\href@noop {} {\bibinfo {title} {{ISODISTORT,
  ISOTROPY Software Suite, iso.byu.edu.}}},\ \bibinfo {howpublished}
  {\url{https://stokes.byu.edu/iso/isodistort.php}} ({\natexlab{b}}),\ \bibinfo
  {note} {accessed: 2023-06-01}\BibitemShut {NoStop}%
\bibitem [{\citenamefont {Campbell}\ \emph {et~al.}(2006)\citenamefont
  {Campbell}, \citenamefont {Stokes}, \citenamefont {Tanner},\ and\
  \citenamefont {Hatch}}]{Campbell2006}%
  \BibitemOpen
  \bibfield  {author} {\bibinfo {author} {\bibfnamefont {B.~J.}\ \bibnamefont
  {Campbell}}, \bibinfo {author} {\bibfnamefont {H.~T.}\ \bibnamefont
  {Stokes}}, \bibinfo {author} {\bibfnamefont {D.~E.}\ \bibnamefont {Tanner}},\
  and\ \bibinfo {author} {\bibfnamefont {D.~M.}\ \bibnamefont {Hatch}},\
  }\bibfield  {title} {\bibinfo {title} {{{\it ISODISPLACE}: a web-based tool
  for exploring structural distortions}},\ }\href
  {https://doi.org/10.1107/S0021889806014075} {\bibfield  {journal} {\bibinfo
  {journal} {Journal of Applied Crystallography}\ }\textbf {\bibinfo {volume}
  {39}},\ \bibinfo {pages} {607} (\bibinfo {year} {2006})}\BibitemShut
  {NoStop}%
\bibitem [{\citenamefont {{H. T. Stokes, D. M. Hatch, and B. J.
  Campbell}}({\natexlab{c}})}]{INVARIANTS}%
  \BibitemOpen
  \bibfield  {author} {\bibinfo {author} {\bibnamefont {{H. T. Stokes, D. M.
  Hatch, and B. J. Campbell}}},\ }\href@noop {} {\bibinfo {title} {{INVARIANTS,
  ISOTROPY Software Suite, iso.byu.edu.}}},\ \bibinfo {howpublished}
  {\url{https://stokes.byu.edu/iso/invariants.php}} ({\natexlab{c}}),\ \bibinfo
  {note} {accessed: 2023-07-01}\BibitemShut {NoStop}%
\bibitem [{\citenamefont {Hatch}\ and\ \citenamefont
  {Stokes}(2003)}]{Hatch2003}%
  \BibitemOpen
  \bibfield  {author} {\bibinfo {author} {\bibfnamefont {D.~M.}\ \bibnamefont
  {Hatch}}\ and\ \bibinfo {author} {\bibfnamefont {H.~T.}\ \bibnamefont
  {Stokes}},\ }\bibfield  {title} {\bibinfo {title} {{{\it INVARIANTS}: program
  for obtaining a list of invariant polynomials of the order-parameter
  components associated with irreducible representations of a space group}},\
  }\href {https://doi.org/10.1107/S0021889803005946} {\bibfield  {journal}
  {\bibinfo  {journal} {Journal of Applied Crystallography}\ }\textbf {\bibinfo
  {volume} {36}},\ \bibinfo {pages} {951} (\bibinfo {year} {2003})}\BibitemShut
  {NoStop}%
\bibitem [{\citenamefont {Aroyo}\ \emph {et~al.}(2011)\citenamefont {Aroyo},
  \citenamefont {Perez-Mato}, \citenamefont {Orobengoa}, \citenamefont {Tasci},
  \citenamefont {de~la Flor},\ and\ \citenamefont {Kirov}}]{Aroyo2011}%
  \BibitemOpen
  \bibfield  {author} {\bibinfo {author} {\bibfnamefont {M.~I.}\ \bibnamefont
  {Aroyo}}, \bibinfo {author} {\bibfnamefont {J.~M.}\ \bibnamefont
  {Perez-Mato}}, \bibinfo {author} {\bibfnamefont {D.}~\bibnamefont
  {Orobengoa}}, \bibinfo {author} {\bibfnamefont {E.}~\bibnamefont {Tasci}},
  \bibinfo {author} {\bibfnamefont {G.}~\bibnamefont {de~la Flor}},\ and\
  \bibinfo {author} {\bibfnamefont {A.}~\bibnamefont {Kirov}},\ }\bibfield
  {title} {\bibinfo {title} {Crystallography online: Bilbao crystallographic
  server},\ }\href@noop {} {\bibfield  {journal} {\bibinfo  {journal} {Bulg.
  Chem. Commun}\ }\textbf {\bibinfo {volume} {43}},\ \bibinfo {pages} {183}
  (\bibinfo {year} {2011})}\BibitemShut {NoStop}%
\bibitem [{\citenamefont {Momma}\ and\ \citenamefont
  {Izumi}(2011)}]{Momma2011}%
  \BibitemOpen
  \bibfield  {author} {\bibinfo {author} {\bibfnamefont {K.}~\bibnamefont
  {Momma}}\ and\ \bibinfo {author} {\bibfnamefont {F.}~\bibnamefont {Izumi}},\
  }\bibfield  {title} {\bibinfo {title} {Vesta 3 for three-dimensional
  visualization of crystal, volumetric and morphology data},\ }\href@noop {}
  {\bibfield  {journal} {\bibinfo  {journal} {Journal of applied
  crystallography}\ }\textbf {\bibinfo {volume} {44}},\ \bibinfo {pages} {1272}
  (\bibinfo {year} {2011})}\BibitemShut {NoStop}%
\bibitem [{\citenamefont {Togo}\ and\ \citenamefont {Tanaka}(2015)}]{Togo2015}%
  \BibitemOpen
  \bibfield  {author} {\bibinfo {author} {\bibfnamefont {A.}~\bibnamefont
  {Togo}}\ and\ \bibinfo {author} {\bibfnamefont {I.}~\bibnamefont {Tanaka}},\
  }\bibfield  {title} {\bibinfo {title} {First principles phonon calculations
  in materials science},\ }\href
  {https://doi.org/https://doi.org/10.1016/j.scriptamat.2015.07.021} {\bibfield
   {journal} {\bibinfo  {journal} {Scripta Materialia}\ }\textbf {\bibinfo
  {volume} {108}},\ \bibinfo {pages} {1} (\bibinfo {year} {2015})}\BibitemShut
  {NoStop}%
\bibitem [{\citenamefont {Monkhorst}\ and\ \citenamefont
  {Pack}(1976)}]{Monkhorst1976}%
  \BibitemOpen
  \bibfield  {author} {\bibinfo {author} {\bibfnamefont {H.~J.}\ \bibnamefont
  {Monkhorst}}\ and\ \bibinfo {author} {\bibfnamefont {J.~D.}\ \bibnamefont
  {Pack}},\ }\bibfield  {title} {\bibinfo {title} {Special points for
  brillouin-zone integrations},\ }\href
  {https://doi.org/10.1103/PhysRevB.13.5188} {\bibfield  {journal} {\bibinfo
  {journal} {Phys. Rev. B}\ }\textbf {\bibinfo {volume} {13}},\ \bibinfo
  {pages} {5188} (\bibinfo {year} {1976})}\BibitemShut {NoStop}%
\bibitem [{\citenamefont {Mulder}\ \emph {et~al.}(2013)\citenamefont {Mulder},
  \citenamefont {Benedek}, \citenamefont {Rondinelli},\ and\ \citenamefont
  {Fennie}}]{Mulder2013}%
  \BibitemOpen
  \bibfield  {author} {\bibinfo {author} {\bibfnamefont {A.~T.}\ \bibnamefont
  {Mulder}}, \bibinfo {author} {\bibfnamefont {N.~A.}\ \bibnamefont {Benedek}},
  \bibinfo {author} {\bibfnamefont {J.~M.}\ \bibnamefont {Rondinelli}},\ and\
  \bibinfo {author} {\bibfnamefont {C.~J.}\ \bibnamefont {Fennie}},\ }\bibfield
   {title} {\bibinfo {title} {{Turning {ABO$_3$} Antiferroelectrics into
  Ferroelectrics: Design Rules for Practical Rotation-Driven Ferroelectricity
  in Double Perovskites and {A$_3$B$_2$O$_7$} Ruddlesden-Popper Compounds}},\
  }\href {https://doi.org/https://doi.org/10.1002/adfm.201300210} {\bibfield
  {journal} {\bibinfo  {journal} {Advanced Functional Materials}\ }\textbf
  {\bibinfo {volume} {23}},\ \bibinfo {pages} {4810} (\bibinfo {year}
  {2013})}\BibitemShut {NoStop}%
\bibitem [{\citenamefont {Li}\ and\ \citenamefont {Birol}(2021)}]{Li2021Free}%
  \BibitemOpen
  \bibfield  {author} {\bibinfo {author} {\bibfnamefont {S.}~\bibnamefont
  {Li}}\ and\ \bibinfo {author} {\bibfnamefont {T.}~\bibnamefont {Birol}},\
  }\bibfield  {title} {\bibinfo {title} {Free-carrier-induced ferroelectricity
  in layered perovskites},\ }\href
  {https://doi.org/10.1103/PhysRevLett.127.087601} {\bibfield  {journal}
  {\bibinfo  {journal} {Phys. Rev. Lett.}\ }\textbf {\bibinfo {volume} {127}},\
  \bibinfo {pages} {087601} (\bibinfo {year} {2021})}\BibitemShut {NoStop}%
\bibitem [{Sup()}]{Supplement}%
  \BibitemOpen
  \href@noop {} {}\bibinfo {note} {See the supplementary information for
  details.}\BibitemShut {Stop}%
\bibitem [{\citenamefont {Zhang}\ \emph {et~al.}(2017)\citenamefont {Zhang},
  \citenamefont {Wang}, \citenamefont {Sahoo}, \citenamefont {Shimada},\ and\
  \citenamefont {Kitamura}}]{Zhang2017Stannate}%
  \BibitemOpen
  \bibfield  {author} {\bibinfo {author} {\bibfnamefont {Y.}~\bibnamefont
  {Zhang}}, \bibinfo {author} {\bibfnamefont {J.}~\bibnamefont {Wang}},
  \bibinfo {author} {\bibfnamefont {M.~P.~K.}\ \bibnamefont {Sahoo}}, \bibinfo
  {author} {\bibfnamefont {T.}~\bibnamefont {Shimada}},\ and\ \bibinfo {author}
  {\bibfnamefont {T.}~\bibnamefont {Kitamura}},\ }\bibfield  {title} {\bibinfo
  {title} {Strain-induced ferroelectricity and lattice coupling in {BaSnO$_3$}
  and {SrSnO$_3$}},\ }\href {https://doi.org/10.1039/C7CP03952B} {\bibfield
  {journal} {\bibinfo  {journal} {Physical Chemistry Chemical Physics}\
  }\textbf {\bibinfo {volume} {19}},\ \bibinfo {pages} {26047} (\bibinfo {year}
  {2017})}\BibitemShut {NoStop}%
\bibitem [{\citenamefont {Bersuker}(1966)}]{Bersuker1966}%
  \BibitemOpen
  \bibfield  {author} {\bibinfo {author} {\bibfnamefont {I.}~\bibnamefont
  {Bersuker}},\ }\bibfield  {title} {\bibinfo {title} {On the origin of
  ferroelectricity in perovskite-type crystals},\ }\href
  {https://doi.org/10.1016/0031-9163(66)91127-9} {\bibfield  {journal}
  {\bibinfo  {journal} {Physics Letters}\ }\textbf {\bibinfo {volume} {20}},\
  \bibinfo {pages} {589} (\bibinfo {year} {1966})}\BibitemShut {NoStop}%
\bibitem [{\citenamefont {Cohen}\ and\ \citenamefont
  {Krakauer}(1990)}]{Cohen1990}%
  \BibitemOpen
  \bibfield  {author} {\bibinfo {author} {\bibfnamefont {R.}~\bibnamefont
  {Cohen}}\ and\ \bibinfo {author} {\bibfnamefont {H.}~\bibnamefont
  {Krakauer}},\ }\bibfield  {title} {\bibinfo {title} {Lattice dynamics and
  origin of ferroelectricity in {BaTiO$_3$}: Linearized-augmented-plane-wave
  total-energy calculations},\ }\href@noop {} {\bibfield  {journal} {\bibinfo
  {journal} {Physical Review B}\ }\textbf {\bibinfo {volume} {42}},\ \bibinfo
  {pages} {6416} (\bibinfo {year} {1990})}\BibitemShut {NoStop}%
\bibitem [{\citenamefont {Benedek}\ and\ \citenamefont
  {Birol}(2016)}]{Benedek2016}%
  \BibitemOpen
  \bibfield  {author} {\bibinfo {author} {\bibfnamefont {N.~A.}\ \bibnamefont
  {Benedek}}\ and\ \bibinfo {author} {\bibfnamefont {T.}~\bibnamefont
  {Birol}},\ }\bibfield  {title} {\bibinfo {title} {‘ferroelectric’ metals
  reexamined: fundamental mechanisms and design considerations for new
  materials},\ }\href {https://doi.org/10.1039/C5TC03856A} {\bibfield
  {journal} {\bibinfo  {journal} {Journal of Materials Chemistry C}\ }\textbf
  {\bibinfo {volume} {4}},\ \bibinfo {pages} {4000} (\bibinfo {year}
  {2016})}\BibitemShut {NoStop}%
\bibitem [{Ami(2012)}]{Amisi2012}%
  \BibitemOpen
  \bibfield  {title} {\bibinfo {title} {First-principles study of structural
  and vibrational properties of {SrZrO$_3$}},\ }\href
  {https://doi.org/10.1103/PhysRevB.85.064112} {\bibfield  {journal} {\bibinfo
  {journal} {Physical Review B}\ }\textbf {\bibinfo {volume} {85}},\ \bibinfo
  {pages} {064112} (\bibinfo {year} {2012})}\BibitemShut {NoStop}%
\bibitem [{\citenamefont {Haeni}\ \emph {et~al.}(2004)\citenamefont {Haeni},
  \citenamefont {Irvin}, \citenamefont {Chang}, \citenamefont {Uecker},
  \citenamefont {Reiche}, \citenamefont {Li}, \citenamefont {Choudhury},
  \citenamefont {Tian}, \citenamefont {Hawley}, \citenamefont {Craigo},
  \citenamefont {Tagantsev}, \citenamefont {Pan}, \citenamefont {Streiffer},
  \citenamefont {Chen}, \citenamefont {Kirchoefer}, \citenamefont {Levy},\ and\
  \citenamefont {Schlom}}]{Haeni2004}%
  \BibitemOpen
  \bibfield  {author} {\bibinfo {author} {\bibfnamefont {J.~H.}\ \bibnamefont
  {Haeni}}, \bibinfo {author} {\bibfnamefont {P.}~\bibnamefont {Irvin}},
  \bibinfo {author} {\bibfnamefont {W.}~\bibnamefont {Chang}}, \bibinfo
  {author} {\bibfnamefont {R.}~\bibnamefont {Uecker}}, \bibinfo {author}
  {\bibfnamefont {P.}~\bibnamefont {Reiche}}, \bibinfo {author} {\bibfnamefont
  {Y.~L.}\ \bibnamefont {Li}}, \bibinfo {author} {\bibfnamefont
  {S.}~\bibnamefont {Choudhury}}, \bibinfo {author} {\bibfnamefont
  {W.}~\bibnamefont {Tian}}, \bibinfo {author} {\bibfnamefont {M.~E.}\
  \bibnamefont {Hawley}}, \bibinfo {author} {\bibfnamefont {B.}~\bibnamefont
  {Craigo}}, \bibinfo {author} {\bibfnamefont {A.~K.}\ \bibnamefont
  {Tagantsev}}, \bibinfo {author} {\bibfnamefont {X.~Q.}\ \bibnamefont {Pan}},
  \bibinfo {author} {\bibfnamefont {S.~K.}\ \bibnamefont {Streiffer}}, \bibinfo
  {author} {\bibfnamefont {L.~Q.}\ \bibnamefont {Chen}}, \bibinfo {author}
  {\bibfnamefont {S.~W.}\ \bibnamefont {Kirchoefer}}, \bibinfo {author}
  {\bibfnamefont {J.}~\bibnamefont {Levy}},\ and\ \bibinfo {author}
  {\bibfnamefont {D.~G.}\ \bibnamefont {Schlom}},\ }\bibfield  {title}
  {\bibinfo {title} {Room-temperature ferroelectricity in strained srtio3},\
  }\href {https://doi.org/10.1038/nature02773} {\bibfield  {journal} {\bibinfo
  {journal} {Nature}\ }\textbf {\bibinfo {volume} {430}},\ \bibinfo {pages}
  {758} (\bibinfo {year} {2004})}\BibitemShut {NoStop}%
\bibitem [{\citenamefont {Fennie}\ and\ \citenamefont
  {Rabe}(2006)}]{Fennie2006ETO}%
  \BibitemOpen
  \bibfield  {author} {\bibinfo {author} {\bibfnamefont {C.~J.}\ \bibnamefont
  {Fennie}}\ and\ \bibinfo {author} {\bibfnamefont {K.~M.}\ \bibnamefont
  {Rabe}},\ }\bibfield  {title} {\bibinfo {title} {Magnetic and electric phase
  control in epitaxial eutio3 from first principles},\ }\href
  {https://doi.org/10.1103/PhysRevLett.97.267602} {\bibfield  {journal}
  {\bibinfo  {journal} {Physical Review Letters}\ }\textbf {\bibinfo {volume}
  {97}},\ \bibinfo {pages} {1} (\bibinfo {year} {2006})}\BibitemShut {NoStop}%
\bibitem [{\citenamefont {Lee}\ \emph {et~al.}(2010)\citenamefont {Lee},
  \citenamefont {Fang}, \citenamefont {Vlahos}, \citenamefont {Ke},
  \citenamefont {Jung}, \citenamefont {Kourkoutis}, \citenamefont {Kim},
  \citenamefont {Ryan}, \citenamefont {Heeg}, \citenamefont {Roeckerath},
  \citenamefont {Goian}, \citenamefont {Bernhagen}, \citenamefont {Uecker},
  \citenamefont {Hammel}, \citenamefont {Rabe}, \citenamefont {Kamba},
  \citenamefont {Schubert}, \citenamefont {Freeland}, \citenamefont {Muller},
  \citenamefont {Fennie}, \citenamefont {Schiffer}, \citenamefont {Gopalan},
  \citenamefont {Johnston-Halperin},\ and\ \citenamefont
  {Schlom}}]{Lee2010ETO}%
  \BibitemOpen
  \bibfield  {author} {\bibinfo {author} {\bibfnamefont {J.~H.}\ \bibnamefont
  {Lee}}, \bibinfo {author} {\bibfnamefont {L.}~\bibnamefont {Fang}}, \bibinfo
  {author} {\bibfnamefont {E.}~\bibnamefont {Vlahos}}, \bibinfo {author}
  {\bibfnamefont {X.}~\bibnamefont {Ke}}, \bibinfo {author} {\bibfnamefont
  {Y.~W.}\ \bibnamefont {Jung}}, \bibinfo {author} {\bibfnamefont {L.~F.}\
  \bibnamefont {Kourkoutis}}, \bibinfo {author} {\bibfnamefont {J.-W.}\
  \bibnamefont {Kim}}, \bibinfo {author} {\bibfnamefont {P.~J.}\ \bibnamefont
  {Ryan}}, \bibinfo {author} {\bibfnamefont {T.}~\bibnamefont {Heeg}}, \bibinfo
  {author} {\bibfnamefont {M.}~\bibnamefont {Roeckerath}}, \bibinfo {author}
  {\bibfnamefont {V.}~\bibnamefont {Goian}}, \bibinfo {author} {\bibfnamefont
  {M.}~\bibnamefont {Bernhagen}}, \bibinfo {author} {\bibfnamefont
  {R.}~\bibnamefont {Uecker}}, \bibinfo {author} {\bibfnamefont {P.~C.}\
  \bibnamefont {Hammel}}, \bibinfo {author} {\bibfnamefont {K.~M.}\
  \bibnamefont {Rabe}}, \bibinfo {author} {\bibfnamefont {S.}~\bibnamefont
  {Kamba}}, \bibinfo {author} {\bibfnamefont {J.}~\bibnamefont {Schubert}},
  \bibinfo {author} {\bibfnamefont {J.~W.}\ \bibnamefont {Freeland}}, \bibinfo
  {author} {\bibfnamefont {D.~A.}\ \bibnamefont {Muller}}, \bibinfo {author}
  {\bibfnamefont {C.~J.}\ \bibnamefont {Fennie}}, \bibinfo {author}
  {\bibfnamefont {P.}~\bibnamefont {Schiffer}}, \bibinfo {author}
  {\bibfnamefont {V.}~\bibnamefont {Gopalan}}, \bibinfo {author} {\bibfnamefont
  {E.}~\bibnamefont {Johnston-Halperin}},\ and\ \bibinfo {author}
  {\bibfnamefont {D.~G.}\ \bibnamefont {Schlom}},\ }\bibfield  {title}
  {\bibinfo {title} {A strong ferroelectric ferromagnet created by means of
  spin–lattice coupling},\ }\href {https://doi.org/10.1038/nature09331}
  {\bibfield  {journal} {\bibinfo  {journal} {Nature}\ }\textbf {\bibinfo
  {volume} {466}},\ \bibinfo {pages} {954} (\bibinfo {year}
  {2010})}\BibitemShut {NoStop}%
\bibitem [{\citenamefont {Birol}\ \emph {et~al.}(2011)\citenamefont {Birol},
  \citenamefont {Benedek},\ and\ \citenamefont {Fennie}}]{Birol2011STO}%
  \BibitemOpen
  \bibfield  {author} {\bibinfo {author} {\bibfnamefont {T.}~\bibnamefont
  {Birol}}, \bibinfo {author} {\bibfnamefont {N.~A.}\ \bibnamefont {Benedek}},\
  and\ \bibinfo {author} {\bibfnamefont {C.~J.}\ \bibnamefont {Fennie}},\
  }\bibfield  {title} {\bibinfo {title} {Interface control of emergent ferroic
  order in ruddlesden-popper srn+1tino3n+1},\ }\href
  {https://doi.org/10.1103/PhysRevLett.107.257602} {\bibfield  {journal}
  {\bibinfo  {journal} {Physical Review Letters}\ }\textbf {\bibinfo {volume}
  {107}},\ \bibinfo {pages} {257602} (\bibinfo {year} {2011})}\BibitemShut
  {NoStop}%
\bibitem [{\citenamefont {Lee}\ \emph {et~al.}(2013)\citenamefont {Lee},
  \citenamefont {Orloff}, \citenamefont {Birol}, \citenamefont {Zhu},
  \citenamefont {Goian}, \citenamefont {Rocas}, \citenamefont {Haislmaier},
  \citenamefont {Vlahos}, \citenamefont {Mundy}, \citenamefont {Kourkoutis},
  \citenamefont {Nie}, \citenamefont {Biegalski}, \citenamefont {Zhang},
  \citenamefont {Bernhagen}, \citenamefont {Benedek}, \citenamefont {Kim},
  \citenamefont {Brock}, \citenamefont {Uecker}, \citenamefont {Xi},
  \citenamefont {Gopalan}, \citenamefont {Nuzhnyy}, \citenamefont {Kamba},
  \citenamefont {Muller}, \citenamefont {Takeuchi}, \citenamefont {Booth},
  \citenamefont {Fennie},\ and\ \citenamefont {Schlom}}]{Lee2013Birol}%
  \BibitemOpen
  \bibfield  {author} {\bibinfo {author} {\bibfnamefont {C.-H.~H.}\
  \bibnamefont {Lee}}, \bibinfo {author} {\bibfnamefont {N.~D.}\ \bibnamefont
  {Orloff}}, \bibinfo {author} {\bibfnamefont {T.}~\bibnamefont {Birol}},
  \bibinfo {author} {\bibfnamefont {Y.}~\bibnamefont {Zhu}}, \bibinfo {author}
  {\bibfnamefont {V.}~\bibnamefont {Goian}}, \bibinfo {author} {\bibfnamefont
  {E.}~\bibnamefont {Rocas}}, \bibinfo {author} {\bibfnamefont
  {R.}~\bibnamefont {Haislmaier}}, \bibinfo {author} {\bibfnamefont
  {E.}~\bibnamefont {Vlahos}}, \bibinfo {author} {\bibfnamefont {J.~A.}\
  \bibnamefont {Mundy}}, \bibinfo {author} {\bibfnamefont {L.~F.}\ \bibnamefont
  {Kourkoutis}}, \bibinfo {author} {\bibfnamefont {Y.}~\bibnamefont {Nie}},
  \bibinfo {author} {\bibfnamefont {M.~D.}\ \bibnamefont {Biegalski}}, \bibinfo
  {author} {\bibfnamefont {J.}~\bibnamefont {Zhang}}, \bibinfo {author}
  {\bibfnamefont {M.}~\bibnamefont {Bernhagen}}, \bibinfo {author}
  {\bibfnamefont {N.~A.}\ \bibnamefont {Benedek}}, \bibinfo {author}
  {\bibfnamefont {Y.}~\bibnamefont {Kim}}, \bibinfo {author} {\bibfnamefont
  {J.~D.}\ \bibnamefont {Brock}}, \bibinfo {author} {\bibfnamefont
  {R.}~\bibnamefont {Uecker}}, \bibinfo {author} {\bibfnamefont {X.~X.}\
  \bibnamefont {Xi}}, \bibinfo {author} {\bibfnamefont {V.}~\bibnamefont
  {Gopalan}}, \bibinfo {author} {\bibfnamefont {D.}~\bibnamefont {Nuzhnyy}},
  \bibinfo {author} {\bibfnamefont {S.}~\bibnamefont {Kamba}}, \bibinfo
  {author} {\bibfnamefont {D.~A.}\ \bibnamefont {Muller}}, \bibinfo {author}
  {\bibfnamefont {I.}~\bibnamefont {Takeuchi}}, \bibinfo {author}
  {\bibfnamefont {J.~C.}\ \bibnamefont {Booth}}, \bibinfo {author}
  {\bibfnamefont {C.~J.}\ \bibnamefont {Fennie}},\ and\ \bibinfo {author}
  {\bibfnamefont {D.~G.}\ \bibnamefont {Schlom}},\ }\bibfield  {title}
  {\bibinfo {title} {Exploiting dimensionality and defect mitigation to create
  tunable microwave dielectrics},\ }\href {https://doi.org/10.1038/nature12582}
  {\bibfield  {journal} {\bibinfo  {journal} {Nature}\ }\textbf {\bibinfo
  {volume} {502}},\ \bibinfo {pages} {532} (\bibinfo {year}
  {2013})}\BibitemShut {NoStop}%
\bibitem [{\citenamefont {Haislmaier}\ \emph {et~al.}(2019)\citenamefont
  {Haislmaier}, \citenamefont {Lu}, \citenamefont {Lapano}, \citenamefont
  {Zhou}, \citenamefont {Alem}, \citenamefont {Sinnott}, \citenamefont
  {Engel-Herbert},\ and\ \citenamefont {Gopalan}}]{Haislmaier2019}%
  \BibitemOpen
  \bibfield  {author} {\bibinfo {author} {\bibfnamefont {R.~C.}\ \bibnamefont
  {Haislmaier}}, \bibinfo {author} {\bibfnamefont {Y.}~\bibnamefont {Lu}},
  \bibinfo {author} {\bibfnamefont {J.}~\bibnamefont {Lapano}}, \bibinfo
  {author} {\bibfnamefont {H.}~\bibnamefont {Zhou}}, \bibinfo {author}
  {\bibfnamefont {N.}~\bibnamefont {Alem}}, \bibinfo {author} {\bibfnamefont
  {S.~B.}\ \bibnamefont {Sinnott}}, \bibinfo {author} {\bibfnamefont
  {R.}~\bibnamefont {Engel-Herbert}},\ and\ \bibinfo {author} {\bibfnamefont
  {V.}~\bibnamefont {Gopalan}},\ }\bibfield  {title} {\bibinfo {title} {Large
  tetragonality and room temperature ferroelectricity in compressively strained
  {CaTiO$_3$} thin films},\ }\bibfield  {journal} {\bibinfo  {journal} {APL
  Materials}\ }\textbf {\bibinfo {volume} {7}},\ \href
  {https://doi.org/10.1063/1.5090798} {10.1063/1.5090798} (\bibinfo {year}
  {2019})\BibitemShut {NoStop}%
\bibitem [{\citenamefont {Eklund}\ \emph {et~al.}(2009)\citenamefont {Eklund},
  \citenamefont {Fennie},\ and\ \citenamefont {Rabe}}]{Eklund2009}%
  \BibitemOpen
  \bibfield  {author} {\bibinfo {author} {\bibfnamefont {C.-J.}\ \bibnamefont
  {Eklund}}, \bibinfo {author} {\bibfnamefont {C.~J.}\ \bibnamefont {Fennie}},\
  and\ \bibinfo {author} {\bibfnamefont {K.~M.}\ \bibnamefont {Rabe}},\
  }\bibfield  {title} {\bibinfo {title} {Strain-induced ferroelectricity in
  orthorhombic {CaTiO$_3$} from first principles},\ }\href
  {https://doi.org/10.1103/PhysRevB.79.220101} {\bibfield  {journal} {\bibinfo
  {journal} {Phys. Rev. B}\ }\textbf {\bibinfo {volume} {79}},\ \bibinfo
  {pages} {220101} (\bibinfo {year} {2009})}\BibitemShut {NoStop}%
\bibitem [{\citenamefont {Reyes-Lillo}\ \emph {et~al.}(2019)\citenamefont
  {Reyes-Lillo}, \citenamefont {Rabe},\ and\ \citenamefont
  {Neaton}}]{Reyes-Lillo2019}%
  \BibitemOpen
  \bibfield  {author} {\bibinfo {author} {\bibfnamefont {S.~E.}\ \bibnamefont
  {Reyes-Lillo}}, \bibinfo {author} {\bibfnamefont {K.~M.}\ \bibnamefont
  {Rabe}},\ and\ \bibinfo {author} {\bibfnamefont {J.~B.}\ \bibnamefont
  {Neaton}},\ }\bibfield  {title} {\bibinfo {title} {{Ferroelectricity in
  [111]-oriented epitaxially strained {SrTiO$_3$} from first principles}},\
  }\href {https://doi.org/10.1103/PhysRevMaterials.3.030601} {\bibfield
  {journal} {\bibinfo  {journal} {Physical Review Materials}\ }\textbf
  {\bibinfo {volume} {3}},\ \bibinfo {pages} {030601} (\bibinfo {year}
  {2019})}\BibitemShut {NoStop}%
\bibitem [{\citenamefont {Oja}\ \emph {et~al.}(2008)\citenamefont {Oja},
  \citenamefont {Johnston}, \citenamefont {Frantti},\ and\ \citenamefont
  {Nieminen}}]{Oja2008}%
  \BibitemOpen
  \bibfield  {author} {\bibinfo {author} {\bibfnamefont {R.}~\bibnamefont
  {Oja}}, \bibinfo {author} {\bibfnamefont {K.}~\bibnamefont {Johnston}},
  \bibinfo {author} {\bibfnamefont {J.}~\bibnamefont {Frantti}},\ and\ \bibinfo
  {author} {\bibfnamefont {R.~M.}\ \bibnamefont {Nieminen}},\ }\bibfield
  {title} {\bibinfo {title} {Computational study of (111) epitaxially strained
  ferroelectric perovskites {BaTiO$_3$} and {PbTiO$_3$}},\ }\href@noop {}
  {\bibfield  {journal} {\bibinfo  {journal} {Physical Review B}\ }\textbf
  {\bibinfo {volume} {78}},\ \bibinfo {pages} {094102} (\bibinfo {year}
  {2008})}\BibitemShut {NoStop}%
\end{thebibliography}
\end{document}